# Beyond Rescorla-Wagner: the Ups and Downs of Learning

**Gianluca Calcagni**[*]

*Instituto de Estructura de la Materia, CSIC, Madrid, Spain*

**Justin A. Harris**

*School of Psychology, University of Sydney, Sydney, Australia*

**Ricardo Pellón**

*Facultad de Psicología, UNED, Madrid, Spain*

**Abstract**

We check the robustness of a recently proposed dynamical model of associative Pavlovian learning that extends the Rescorla-Wagner (RW) model in a natural way and predicts progressively damped oscillations in the response of the subjects. Using the data of two experiments, we compare the dynamical oscillatory model (DOM) with an oscillatory model made of the superposition of the RW learning curve and oscillations. Not only do data clearly show an oscillatory pattern, but they also favor the DOM over the added oscillation model, thus pointing out that these oscillations are the manifestation of an associative process. The latter is interpreted as the fact that subjects make predictions on trial outcomes more extended in time than in the RW model, but with more uncertainty.

## 1. Introduction

Empirical data of Pavlovian conditioning experiments sometimes show a curious oscillatory pattern where the subject response fluctuates before reaching the asymptote of learning. Clear examples of response fluctuations are shown in Fig. 6 of Ghirlanda and Ibadullaiev (2015), which reproduces, among others, results of Miller et al. (1981) and Zelikowsky and Fanselow (2010). In particular, the data from Miller et al. (1981) show fluctuations in response rates above and below the idealized acquisition curve, while the data from Zelikowsky and Fanselow (2010) show a clear overshoot in initial responding that is corrected in later trials. Other examples of long-range response fluctuations, taken from the experiments we will discuss in this paper, are shown in Fig. 1.

---

[*] Corresponding author: g.calcagni@csic.es.

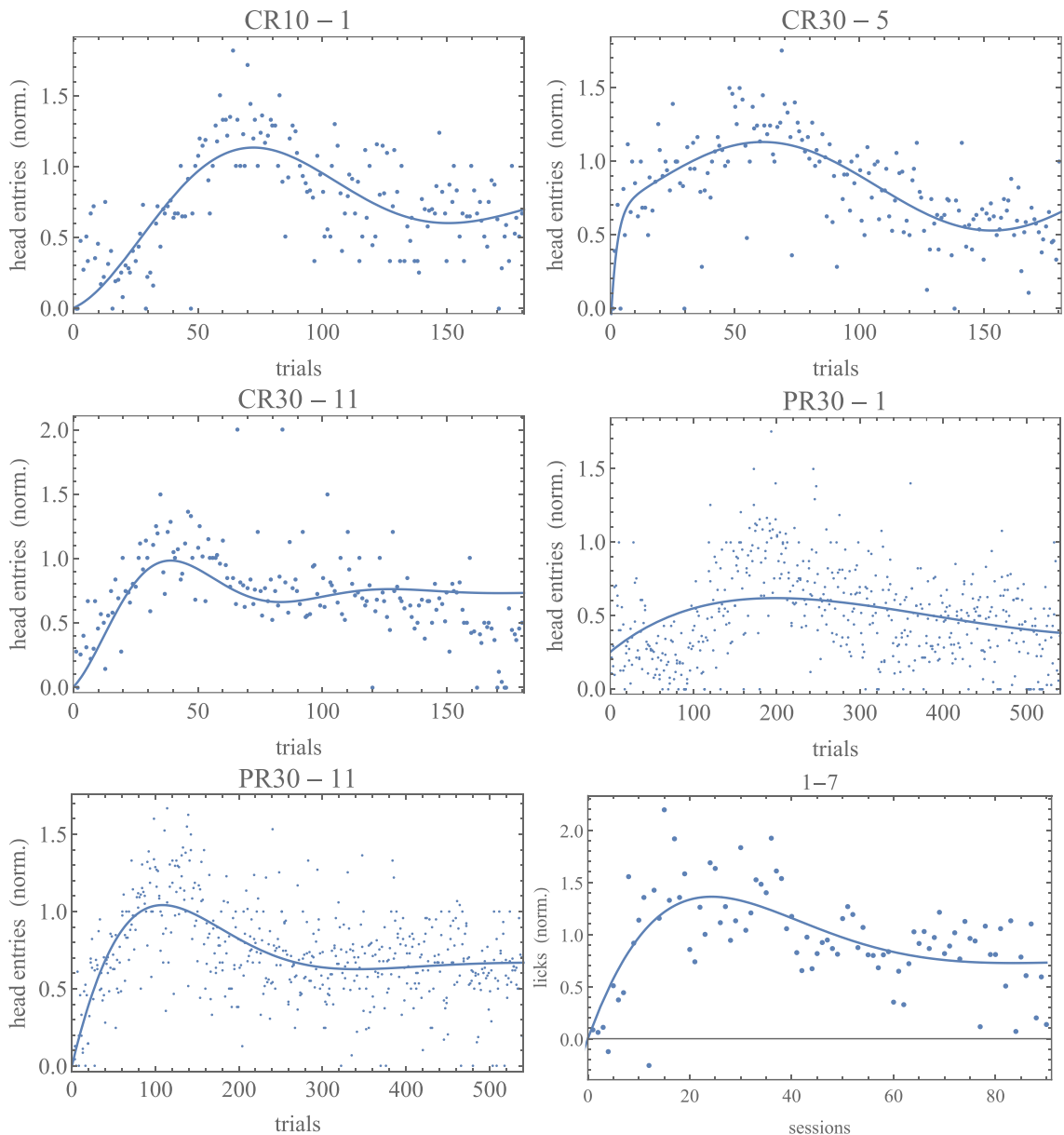

*Figure 1: Nonmonotonic learning curves of individual subjects of an experiment of Harris et al. (2015) (subjects CR10-1, CR30-5, CR30-11, PR30-1, PR30-11) and of Calcagni et al. (2020) (subject 1-7), described more in detail in the text. Note that trial data are neither binned nor averaged.*

These are not the only available examples and, in fact, oscillations (not to be confused with the shortest-scale phenomenon of post-peak depression; for a discussion, see Calcagni, Caballero-Garrido and Pellón, 2020) may be more common than currently acknowledged, partly because there are not looked for or are ignored as experimental errors. In turn, they might not be looked for because, to begin with, the existing theories do not predict their occurrence. For instance, the Rescorla–Wagner (RW) model (Rescorla and Wagner, 1972; Wagner and Rescorla, 1972; Wagner and Vogel, 2009), one of the simplest available quantitative descriptions of the learning curve, does not show oscillations. In the present work, we propose and verify a model that extends RW to cover fluctuating responses.

The RW model opened up a new way of thinking about associative Pavlovian learning. While it was not the first mathematical model of learning (Bush and Mosteller, 1951a; 1951b; Estes, 1950), the RW model shifted the focus from response probability to association strength and thereby opened the way to explain new conditioning phenomena. Like any mathematical model, it has limitations (Miller, Barnet, and Grahame, 1995), which spurred further development of associative learning theory (e.g., Ghirlanda and Enqvist, 2019; Le Pelley, 2004; Mackintosh, 1975; Pearce and Hall, 1980; Wagner, 1981; Wagner and Vogel, 2009). Here, we will extend the RW model to consider oscillations in responding, as observed in several studies (Calcagni et al., 2020; Harris, Patterson and Gharaei, 2015; Miller et al., 1981; Zelikowsky and Fanselow, 2010). Because the relevant property of the RW model is the same for both the single-cue and multiple-cue versions of the model, we will focus on the single-cue version and refer to this as RW.

The role of individual differences and response variations in conditioning models has long been of concern (Hayes, 1953; Merrill, 1931; Sidman, 1952, for initial publications; Blanco and Moris, 2018; Calcagni et al., 2020; Gallistel, 2012; Gallistel, Fairhurst, and Balsam, 2004; Glautier, 2013; Jaksic et al., 2018; Mazur and Hastie, 1978; Smith and Little, 2018; Young, 2018, for recent ones). These differences have provoked a range of attitudes as diverse as discarding them as statistical errors around a clean theoretical learning curve, or taking them as proof that theoretical models are mere artifacts. The textbook monotonic learning curve reaching a stable asymptote (RW model) is clearly an abstraction, but we believe not to the point that we should throw it away without a critical analysis of the problem. The question we pose here is twofold: (1) Can we include individual differences and response fluctuations into a theoretical predictive model? (2) Supposing such a model is obtained, what is its psychological interpretation?

In this article, we use data from different experiments to test one such candidate, proposed in Calcagni et al. (2020): A Pavlovian associative dynamical oscillatory model (henceforth DOM) where the subject's response fluctuates periodically above and below the ideal asymptote of learning until these *long-range oscillations* (spanning several tens of trials) eventually disappear.[1] As discussed in greater detail in Calcagni et al. (2020), this oscillatory pattern is a strongly subject-dependent long-range phenomenon. This means that it is only manifest across a large number of trials and that averaging data across trials can conceal any oscillatory pattern as well as other individual differences. The great majority of published experiments do not encompass a sufficient number of trials and/or the related data analysis averages over the individuals. This suggests that, if oscillations are more common than currently acknowledged, they might have been overlooked simply because of the choice of experimental design or data analysis. The purpose of the present paper is to use two sufficiently long experiments with non-averaged data to unravel this under-recognized phenomenon. Using the Bayesian and Akaike information criteria in a standard model selection procedure, we reanalyze data presented in Harris et al. (2015) and Calcagni et al. (2020) and compare the DOM with the RW model as well as with an oscillating model constructed as an *ad hoc* modification of the RW model designed to mimic response oscillations by adding an oscillatory component that is not governed by an associative process (AOM, added oscillation model). We will find up to very strong evidence that the DOM can fit more individual data than the RW model and the AOM. This analysis will significantly extend and confirm the findings of Calcagni et al. (2020) with more data, more statistics, and with the additional, robust cross-check represented by the AOM.

---

[1] Throughout this article, we differentiate between fluctuations and oscillations. Fluctuations are random deviations from the ideal learning curve that occur on a trial-by-trial time scale. On the contrary, oscillations take place on a much longer range and give rise (ideally) to a smooth, differentiable pattern.

## 2. Four Models of Pavlovian Conditioning

### 2.1 Rescorla–Wagner Model

This simple associative model of Pavlovian conditioning was the result of early efforts in the discipline by researchers such as Hull (1943), Estes (1950), Bush and Mosteller (1951a), and Rescorla and Wagner (1972). The main feature of the RW model is the prediction of the animal's response in the next trial from its response in the last one. Consider an experiment where a conditioned stimulus (CS) with salience $\alpha$ is associated to an unconditioned stimulus (US) with salience $\beta$. The finite difference or increment $\Delta v_n := v_n - v_{n-1}$ of the association between the CS and the US from trial $n-1$ to trial $n$ is proportional to the last prediction error made by the subject, i.e., the difference between $v_{n-1}$ and the optimal learning asymptote $\lambda$:

$$\Delta v_n = \alpha\beta(\lambda - v_{n-1})\,,\tag{1}$$

where $n = 1,2,3,\ldots$ . Therefore, at each trial the association strength is updated on the current value of $v$ and the value of $\lambda$. Equation (1) corresponds to the RW model with a single cue (Rescorla and Wagner, 1972) and its solution (e.g., Calcagni, 2018) is as follows:

$$v_n = \lambda[1 - (1 - \alpha\beta)^{n-1}]\,.\tag{2}$$

The learning sequence Eq. (2) is plotted in Fig. 2 for about 50 trials.

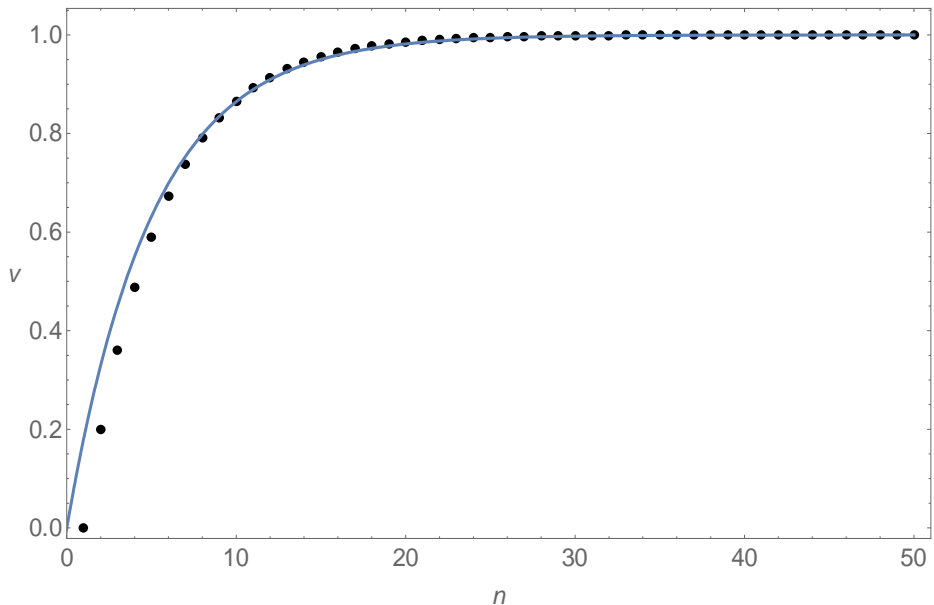

*Figure 2: The RW learning sequence Eq. (2) (black dots) and its continuum approximation, the learning curve Eq. (4) (solid curve), with $\alpha\beta = 0.2$ and $\lambda = 1$.*

Equation (1) is a recursive discrete formula defining the RW model. With many trials, however, we can use a continuum approximation where the progression of learning is measured on a continuous time parameter $t$ instead of a discrete set of trials $n$. Then, the incremental law Eq. (1) becomes as follows:

$$\dot{v} + \alpha\beta(v - \lambda) = 0 \,, \tag{3}$$

where $\dot{v} = dv/dt$ is the first time derivative of the association strength, the infinitesimal approximation of the finite difference $\Delta v_n/\Delta n = \Delta v_n$. Increments or derivatives can be of first or higher order. The RW model has simple incremental steps, corresponding to first-order derivatives. Later, we will see that the DOM happens to have double incremental steps, corresponding to second-order derivatives. As a matter of nomenclature, we recall that Eq. (1) is a so-called recursive or finite-difference equation, while Eq. (3) is a so-called first-order differential equation. Technically, one can always map a discrete (finite-difference) equation into a continuous (differential) one, independently of the order of the incremental steps or derivatives. In general, however, one is entitled to transform a discrete equation into a continuum one when there are enough points and these points are sufficiently packed together, so that an incremental step is small enough. In other words, the continuum approximation is especially good when the increment Eq. (1) is small, i.e., near the asymptote. We can see this in Fig. 2, where the solid curve is the solution of Eq. (3):

$$v(t) = \lambda \left(1 - e^{-\alpha\beta\, t}\right). \tag{4}$$

This curve perfectly overlaps with the data points after only a few trials. One can check that the number of trials needed to achieve almost exact overlap ranges from about 20-30 trials for $\alpha\beta = 0.01$ to just one trial for $\alpha\beta = 0.99$. Since $0 < \alpha\beta < 1$ is the natural range of this product of parameters, we can conclude that the continuum approximation breaks down only when the data set is scarcely populated.

The continuum approximation will be relevant when comparing experimental data with the theoretical model. Best fits are usually presented as continuous curves, even when they stem from a discrete model (Hull, 1943; Spence, 1956). In our case, this representation of the RW, the DOM and the AOM will be legitimate both mathematically and numerically. Technically, the RW model is a sequence of approximations of $\lambda$ generated by each trial. These approximations are points that lie on a curve, but the model is not filling in the line between the points. This would be apparent in any experiment with no more than 10 to 20 points. However, when one fits the model with dozens of points, the difference is so small that one can ignore it to all purposes. In our case, we will deal with a minimum of 90 points to a maximum of 540, so that this difference will be negligible with respect to the precision achieved by a Bayesian model selection in discriminating between different models.

Finally, one may object that the data points fluctuate a lot for typical subjects, which means that their increments $\Delta v_n$ may be not much smaller than 1. In this case, one would expect the continuum approximation to be violated. However, we can employ various tools of analysis to show *a posteriori* that this does not happen. First of all, the phenomenon of random trial-by-trial fluctuations goes well beyond the capabilities of the RW or similar models predicting a smooth learning curve. The oscillations of the DOM and of the AOM are always long-range, i.e., they span several trials, while behavioral noise (i.e., random fluctuations in the response of the subject) characterizes much shorter scales. From the point of view of associative models such as Rescorla–Wagner (1972), Mackintosh (1975), and Pearce–Hall (1980), these fluctuations should be treated as errors. The spectral analysis is a better tool than the learning curve to describe a randomly fluctuating response (Calcagni et al., 2020). At any rate, the dispersion of the data points in the experiments we ran was smaller than the asymptote of learning for all subjects, although not much smaller. This means that response fluctuations from one trial to the next were not excessively large.

## 2.2 Dynamical oscillatory model (DOM)

The RW model Eq. (1) stems from the assumption that the subject learning an association between stimuli does so according to the surprisingness or novelty of the outcome, represented by the difference between the asymptote and the current association strength. The less surprising the appearance of the US is in contiguity with the CS, the smaller the increase in association strength. It is possible to recast this model in very different terms and to regard it as a *dynamical* system in the most rigorous meaning of the word. This reformulation shows that the RW model is a special case of a more general model in which behavioral oscillations are common. The pay-off in doing so is a first principle governing these and any other associative models based on recursive relations.

We are talking about the *principle of least action*, well known in physics but recently adapted to the psychological core of Pavlovian conditioning (Calcagni et al., 2020). Independently of whether time is continuous or discrete, the principle of least action postulates that there is a quantity that is minimized during the evolution of the system. This quantity, the action, represents the efficiency with which an organism (or its neural network) adapts to a new situation such as a learning process. A "biological" interpretation of the action might go along the following lines. At the level of brain structure, learning involves the modification of synaptic connections in a dendritic arborization process through the rearrangement, pruning, and creation of new synapses or dendrites. Due to the huge number of connections, it would be useful to describe this process in terms of emergent or global degrees of freedom. The assumption here is that, globally, the learning process is efficient, despite errors in the performance of the subject. This postulate translates into assuming that the organism minimizes (albeit imperfectly) the "energy" spent in the dendritic arborization or synaptic adjustments that take place during learning. Quantifying this "energy" may be difficult, but there is an alternative. The construct of energy, as used here, is directly related to the construct of "action." Thus, the efficiency postulate leads us to the principle of least action, which describes emergent or global degrees of freedom that are purely behavioral (the association strength is the main one).

The degree of freedom of the action is the association strength $v$, and the requirement that the action be minimized by the dynamical evolution of the system determines the so-called equation of motion for $v$. It is relatively easy to write down the action for $v$ in such a way that the equation of motion (1), or Eq. (3) in a continuum setting, for the RW model be recovered. In other words, the action $S[v]$ is an expression that tells one how a system evolves. Given an action,[2] one imposes the least action principle, stating that $S$ does not change (is stationary) under small variations of $v$. One represents this as the functional variation (also called equation of motion) $\delta S / \delta v = 0$. The incremental law (1) or its analog (3) is the equation of motion derived from a simple, dynamically motivated *Ansatz* for the action.

---

[2] The actions for the RW model and the DOM can be found in section 3.1 of the Supplementary Material of Calcagni et al. (2020). We omit them here to reduce the technical bulk of the presentation. See https://www.frontiersin.org/articles/10.3389/fpsyg.2020.00612/full#h13 for details.

A major drawback of the RW model is that it assumes, in a rather minimalistic fashion, that the subject bases its prediction on what will happen in the ongoing trial only on what they "remember" of the previous one (incremental law (1)). A more general situation would allow the subject to base their behavior on more information than the one admitted in the RW model. We can quantify this issue with two dynamical, mutually equivalent analogies.

The first is to regard the RW as a spring system. The RW model describes the accumulation of a property, the associative strength ($v$), across successive experiences with the US, up to a maximum value. In this sense, it is like a spring under a load and, in fact, the RW equation is a special case of a dynamic system that describes the motion of a spring. However, in a natural setting, one would expect the spring to oscillate to equilibrium instead of reaching it monotonically without any recoil. For a load on a spring to avoid oscillations would require that some other external force gradually releases the load onto the spring with infinitely precise measure.

The second analogy is the one of a ball in a potential well. Assume the continuum approximation for simplicity (nothing relevant changes in the discrete case). The solution $v(t)$ (given by Eq. (4)) can be interpreted as the position of a "particle" or a small ball as it evolves in time when it rolls in a parabolic well $U(v) = (\alpha\beta)^2 (v - \lambda)^2 / 2$. The system is dissipative: this potential well has a rough surface that induces a friction force on the ball. In the RW model, the friction is tuned so precisely that the ball stops *exactly* at the bottom of the well. In other words, we place the ball on one of the slopes of the well, and we let it start roll down simply by a gravitational force pointing downwards. As it rolls down, the ball increases momentum but, at the same time, the friction force makes it decelerate until it magically stops at the minimum asymptotically in the future. Friction is a property of the medium, while momentum is a property of the ball, and there is no reason why these two qualitatively different properties should be related to each other. However, in the RW model they are. The reason is that the coefficient $(\alpha\beta)^2$ in the potential $U(v)$ matches exactly the magnitude of the friction force. All that has been said is valid also in a setting where time is discrete (trials $n$). From one trial to the next, the ball acquires momentum $\Delta v_n$ and it asymptotically reaches the bottom of the potential monotonically. The only difference with respect to the continuum case is that, between one hop and the next, momentum is "frozen" like any other dynamical observable.[3]

Regardless of whether we work in discrete or continuous time, this is not the most natural and most generic situation that a ball in a parabolic well may experience. When there is a numerical mismatch between friction and potential energy, then the ball will roll down to the bottom and climbs up the opposite slope up to a lesser height, from where it rolls back down in a sequence of damped oscillations, eventually resting at the bottom of the well. This is the origin of the DOM: We allow for a mismatch, encoded in a new parameter $\mu$, between friction and ball momentum and the result is an oscillatory pattern dying up at the bottom. The only but crucial difference between RW and the DOM is the presence of one extra parameter in the potential,

$$U(v) = \frac{(\alpha\beta)^2 + \mu^2}{2}(v - \lambda)^2,\tag{5}$$

introduced simply because it is the most general thing one could do to define the action. Just like for the other parameters, the theory does not tell us what the value of $\mu$ is for a given subject; but observations do. We call this line of reasoning sustaining the DOM the *naturalness argument*, to distinguish it from two more arguments we will offer shortly.

---

[3] The idea of adding a friction term to the dynamics and to use the ball-in-the-well analogy is not an entirely novel feature of the DOM. A gradient term is often added to the backpropagation learning equation in artificial neural networks (Ghirlanda and Enquist, 1998; Goodfellow, Bengio and Courville, 2016; McClelland and Rumelhart, 1988). This addition improves the efficiency of the search of the minimum of the error function for a given learning task. In this context, which is very different from ours because it does not aim to fit learning data, the analogy of the ball in a narrow valley is used.

All of this holds for a system evolving with either a continuous or a discrete time parameter. Mathematically and conceptually, there is no difficulty in adopting either view. In the discrete case, the dynamics of the ball is observed as through a sequence of photographic snapshots, while in the continuum case this sequence is so fast that it becomes a movie. However, since the model is defined on a discretum, we first quote its equation of motion as a recursive formula, the DOM counterpart of the incremental law (1):

$$\Delta v_{n+1} - (1 - 2\alpha\beta)\Delta v_n = (\alpha^2\beta^2 + \mu^2)(\lambda - v_{n-1})\,.\tag{6}$$

The right side of this equation reduces to the RW model when $\mu = 0$ (Calcagni et al., 2020). Notice the presence of two increments, one from $n$ to $n + 1$ and another from $n - 1$ to $n$. While the last equation can be reverted to expression (1) with only one increment, one cannot make this simplification in the case of the DOM. Therefore, the learning law (6) of the DOM necessarily involves both increments $\Delta v_{n+1}$ and $\Delta v_n$. While in the RW model the subject can predict the outcome of the next trial (future value $v_{n+1}$) just from the present outcome (present value $v_n$), the DOM needs a longer memory of past states and also the knowledge of the past trial (value $v_{n-1}$) is required. In other words, while in the RW model learning (i.e., the momentum $\Delta v_n$) is proportional to the prediction error $(\lambda - v_{n-1})$, in the DOM the prediction error depends mainly on the learning at the next-to-last trial and, to a lesser degree (because $1 - 2\alpha\beta < 1$), on the learning at the last trial. However, prediction errors weigh more on learning than in the RW model, since the right-hand side of Eq. (6) is augmented by a factor $\mu^2$. Therefore, subjects can learn further in the future what is going to happen, but they do so with greater uncertainty. This is the origin of the overshooting of optimal response and the subsequent readjustments through an oscillatory pattern. We dub this justification of the DOM the *errors-in-learning argument*, and it is perhaps the most compelling one from a psychological point of view. It makes the learning process considerably more flexible, or more realistic, than in the RW model.

Finally, notice that the fact that DOM subjects project their predictions further in the future implies that they will take longer to correct any prediction error, since they will sample more trials before implementing any significant change of behavior. This results in a long-range rather than a mid-range effect, governed by the multi-incremental law (6). In particular, when the frequency $\mu$ is small and the DOM is only a small deviation from the RW model, the process can take a great number of trials.

The infinitesimal version of the combination $\Delta v_{n+1} - \Delta v_n$ is the second-order time derivative $\ddot{v}$, so that the continuum limit of Eq. (6) is as follows:

$$\ddot{v} + 2\alpha\beta\dot{v} + (\alpha^2\beta^2 + \mu^2)(v - \lambda) = 0\,,\tag{7}$$

whose general solution is

$$v(t) = \lambda \left[ 1 - e^{-\alpha\beta\, t}(\cos\mu t + A \sin\mu t) \right], \tag{8}$$

where $A$ is a constant amplitude. An alternative way to parameterize the solution (8) is to write only a cosine term with a phase, $\cos(\mu t + \phi)$, but for data analysis this form would be inconvenient. Also, in front of the cosine the amplitude is fixed to 1 because we want $v(0) = 0$. Therefore, (8) is the most general oscillatory solution compatible with our initial conditions.

The learning profile Eq. (8) has a total of five free parameters, while the RW profile (4) has only three. This leads to a second way to express the naturalness argument justifying the DOM, which we may call the *fine-tuning argument*. For definiteness, we can illustrate it with the continuum version of the models; it is simpler and nonrestrictive.

By definition, the initial conditions for a system described by a second-order differential equation are the value of $v$ and its first time derivative at the initial time $t = 0$ (trial number 1). The RW model can be regarded as a second-order system, i.e., the limit $\mu \to 0$ of the DOM equation (7). In fact, taking the derivative of Eq. (3) with respect to time, one has

$$\ddot{v} + 2\alpha\beta\dot{v} + \alpha^2\beta^2(v - \lambda) = 0. \tag{9}$$

This is exactly equivalent to the first-order equation (3), which means that one of the two initial conditions has become redundant inasmuch as it overdetermines the dynamics of the model. The most common dynamical systems (all endowed with an action) have equations of motions with second-order derivatives and one must specify two initial conditions (position and velocity). The DOM is one such system and is quite unremarkable. Also RW (which also follows the least action principle) looks like a second-order system, but its dynamics can be reduced to a first-order system Eq. (3), something quite anomalous and due to the fact that we fine-tuned the dynamics by setting the parameter $\mu$ exactly equal to zero. Thus, in the RW case, one needs only one initial condition (the position). This is just another way to describe the contrast between the two-state memory in the DOM (retrodiction of two past states from the knowledge of the present one) and the one-state memory in the RW model.

To put it in formulae, call $v_{\mathrm{RW}}(\lambda, \alpha, \beta; t)$ the most general solution (4) of the RW model (9). The initial conditions for an excitatory process are as follows:

$$v_{\mathrm{RW}}(\lambda, \alpha, \beta; 0) = 0, \tag{10a}$$

$$\dot{v}_{\mathrm{RW}}(\lambda, \alpha, \beta; 0) = \lambda\alpha\beta. \tag{10b}$$

The subject starts with zero association strength and a positive incremental rate $\lambda\alpha\beta$. While (10a) tells where the ball starts inside the potential well, (10b) tells with what initial velocity the ball starts to roll down (the subject starts to learn). Now consider the DOM solution (8), denoted as $v_{\mathrm{DOM}}(\lambda, \alpha, \beta, \mu, A; t)$. Since $\cos 0 = 1$ and $\sin 0 = 0$, the initial conditions are as follows:

$$v_{\mathrm{DOM}}(\lambda, \alpha, \beta, \mu, A; 0) = 0, \tag{11a}$$

$$\dot{v}_{\mathrm{DOM}}(\lambda, \alpha, \beta, \mu, A; 0) = \lambda\,(\alpha\beta - \mu\,A). \tag{11b}$$

The catch is that the RW initial condition $\dot{v}_{\mathrm{RW}}(\lambda, \alpha, \beta; 0) = \dot{v}_{\mathrm{DOM}}(\lambda, \alpha, \beta, 0, A; 0)$ is a special case of (11b) achieved by setting $\mu = 0$, so that the ball goes to the bottom of the well monotonically. If one deviates, even infinitesimally, from Eq. (10b), then one does not obtain Eq. (4) as a solution. Theoretically, there is no reason why one should set to zero a parameter that, in general, will take a non-vanishing value. To summarize, the RW model pretends to match a dynamical feature (the initial velocity of the ball) with an environmental one (the friction of the basin surface) without explaining why we forced the system to such a special situation. This fine-tuning problem is, in general, a powerful killer of dynamical models.

Note that we have implicitly given two equivalent ways to describe the overshoot of the minimum by the ball. One, used in Calcagni et al. (2020) and in Eq. (5), compares two models at the level of the equations of motion and states that the ball in the DOM rolls down a potential well with a steeper slope than for the well in the RW model. The other way compares the two models at the level of the initial conditions and states that the potential well is the same in both models (by a rescaling $\alpha\beta \to \sqrt{(\alpha\beta)^2 + \mu^2}$ in Eq. (5)) but, by virtue of the same rescaling, the initial momentum Eq. (10b) of the ball in the DOM is greater than the initial momentum in the RW case. Thus, in the DOM, the ball cannot brake at the bottom and it undergoes damped oscillations.

For completeness, we mention a caveat about oscillations in the RW. Although they are not present in the single-cue case, when more cues are present response fluctuations can take place. In the simplest case, this could happen when discriminating between the context and the compound made of the context and the CS, a two-dimensional system that can in principle sustain oscillations. In general, therefore, a multi-dimensional first-order system is a potential alternative to the one-dimensional second-order DOM. Whether oscillations of multi-cue origin occur with realistic parameter values is an open question, which we will not address in our single-cue settings.

To conclude this subsection, we gave three arguments for the DOM, all based upon the least action principle: the naturalness argument, the errors-in-learning argument, and the fine-tuning argument. The first and the last are equivalent and rely on reasoning in terms of the dynamics. The second argument invokes a more flexible learning process and gives a psychological interpretation of the DOM. Another interpretation based on the efficiency of learning was also given. In Calcagni et al. (2020), only the naturalness argument and the efficiency interpretation were provided.

Ultimately empirical evidence is what matters. Many subjects in the experiments we will describe did show a learning curve with a nonzero $\mu$.

### 2.3 Added Oscillations Model (AOM)

Assume temporarily to discover that the majority of individual data in experiments of Pavlovian conditioning display smooth, long-range oscillations in the response and that these data are well fitted by the DOM (which in our case occurs, as it will be shown later). One may wonder whether this is evidence in favor of the DOM or just of the presence of oscillations that, after all, could be implemented in many different ways.

Clearly, one cannot get oscillations from the RW model. Changing the value of the rate parameter $\alpha\beta$ or the asymptote $\lambda$ only modify, respectively, the slope and final height of the learning curve, not its shape. If, on the other hand, one tries to change the rate parameter on a trial-by-trial basis, then one hits the Mackintosh model (1975), where, however, the salience varies monotonically. To get oscillations, we must introduce at least one new parameter, which is the frequency $\mu$ of the oscillations. It is a quantitative change (setting to nonzero something which is zero for the RW model) that gives rise to a qualitative change (oscillations versus monotonic evolution). The DOM is a model doing just that. Is it the only possibility?

To act as the devil's advocate, let us suppose that the oscillations have nothing to do with learning but reflect some other periodic system. For example, they could be fluctuations in the subject level of hunger or some other motivational factor that is influencing response level but is not part of the learning process. We can conceive an alternative oscillatory model (the AOM) where oscillations are independent of the learning process:

$$v(t) = \lambda\left(1 - e^{-\alpha\beta\,t}\right) + A\sin\mu t + B\cos\mu t\,. \tag{12}$$

In Eq. (12), oscillations are simply added to the RW learning curve, while in the DOM (8), they multiply it as a modulation factor. In Appendix A, we show that this model is *non-associative*[4] because, although it can be derived from an action, it hides an extra degree of freedom (which we call $y$) independent of the associative strength $v$ that has the peculiar property, illustrated by equation (A4), of increasing indefinitely in the future and, therefore, has no interpretation in terms of a learning process with an asymptote. In other words, since $y$ grows exponentially as time goes by, it never saturates at an asymptote. This contrasts with learning processes where the subject's performance reaches an upper limit. Therefore, it is difficult to understand $y$ as a variable related to an associative learning process.

For this reason, the AOM encodes both learning (associative) and motivational (non-associative) elements, both of which can affect the subject's performance. Therefore, the difference between the AOM and the DOM is that the DOM assumes that oscillatory behavior is an intrinsic part of the learning process, whereas the AOM also allows for oscillations but treats those as being a part of performance rather than learning. From a mathematical standpoint, the DOM is simpler, but it requires a significant rethinking of what the learning process is. The AOM is more complex mathematically, but easier to integrate with our existing theories of learning. For that reason, most people working in the field might be more accepting of the AOM and might want to know which of the two models explain data best.

The rationale behind the AOM is the following. Comparing the DOM with the RW model, we will find an overall advantage for the DOM in terms of predictive power. The question, then, is whether this advantage is just because the extra parameters allow the DOM to track fluctuations in responding that cannot be explained by RW. In terms of explaining performance, the DOM fluctuations are probably meaningful, but in terms of the acquisition process they might as well be considered as noise. To check if oscillations are an intrinsic part of acquisition (as in the DOM) or just independent fluctuations, we compare the DOM to the AOM. The AOM is RW plus some ability to track oscillations in performance. Therefore, if the DOM sometimes beat RW just because it can track spurious oscillatory features, then replacing RW with the AOM should result in the opposite trend where the AOM beats the DOM. For the DOM, in fact, oscillations are more constrained since they are a part of the acquisition process, and they are dampened down with continued training, whereas the AOM is better able to explain response fluctuations that have nothing to do with acquisition per se.

---

[4] In the sense of being, on one hand, possibly related to motivation and, on the other hand, in no way related to the associative strength. Therefore, the AOM does not deal with habituation, sensitization, pseudo-conditioning, and other phenomena traditionally categorized as "non-associative."

We will find the opposite trend, the DOM more favored than the AOM in a large number of cases. This will eventually rule out non-associative factors as the principal determinants of the oscillations. In the spirit of Bayesian model selection, we will compare all models among one another instead of only the pairs RW-DOM and DOM-AOM, regardless of their theoretical justification.

Finally, one could object that the AOM is not an effective "control" model because oscillations are not damped, and one would not expect a model not leading to a stable performance to be a better fit than the DOM (where oscillations are naturally damped) for most data. While this issue may not be easily dismissed for exceptionally long experiments such as the one of Calcagni et al. (2020), it has little or no impact in all other typical cases where the period of the oscillations is comparable with the time scale of the learning curve, as one can verify *a posteriori* by looking at data (including those of Harris et al., 2015). Therefore, since most of the data we use come from an experiment where the total number of trials is of order of the fluctuations' wavelength, cases where the AOM is a better fit than the DOM are not artifacts.

## 3. Two Experiments

We now introduce two experiments of Pavlovian conditioning with which to test the theoretical models of the previous section. A full, more detailed description of these experiments can be found in the original references.

In Experiment 2 of Harris et al. (2015), 16 male Hooded Wistar rats were used, with unrestricted access to water and restricted daily food rations. The US was food delivered in a dispenser (head entries were recorded) and four different CSs were used: white noise, a tone, a flashing light, and a steady light. For each rat, each CS was allocated to one of the following configurations (and counterbalanced across rats): CR10: continuous reinforcement (US presented at 100% of the trials) with CS random duration of 10 s mean, 30 sessions, 6 trials per session; CR30: continuous reinforcement with CS random duration of 30 s mean, 30 sessions, 6 trials per session; PR10: partial reinforcement (US presented at 33% of the trials) with CS random duration of 10 s mean, 30 sessions, 18 trials per session; PR30: partial reinforcement with CS random duration of 30 s mean, 30 sessions, 18 trials per session. Trial by trial, the CS duration varied randomly on a uniform distribution with a mean of either 10 s or 30 s, according to the name of the group. The number of reinforced trials per session per CS was the same in all configurations and equal to 6. Each of the 30 sessions consisted in a delayed conditioning where presentations of each of the four CSs were randomly intermixed: 6 of each of the continuously reinforced CSs and 18 of each of the partially reinforced CSs, for a total of 48 trials per session.

In the experiment of Calcagni et al. (2020), we employed 32 male Wistar Han rats, without food or water restriction, divided into two experimental groups (US: saccharine solution at 0.1% - Group 1- and 0.2% concentration - Group 2) and two control groups. One experimental subject was removed due to poor health. The US was delivered in individual conditioning boxes via a water pump activated by a solenoid valve. In the case of experimental subjects, delivery happened on a random-time 5 s schedule (RT-5) implemented as a uniform random distribution during the presentation of a 10-s tone (CS). Licks were automatically recorded. We took 90 sessions each comprising 44 trials.

The task diagrams for these two experiments are shown in Fig. 3.

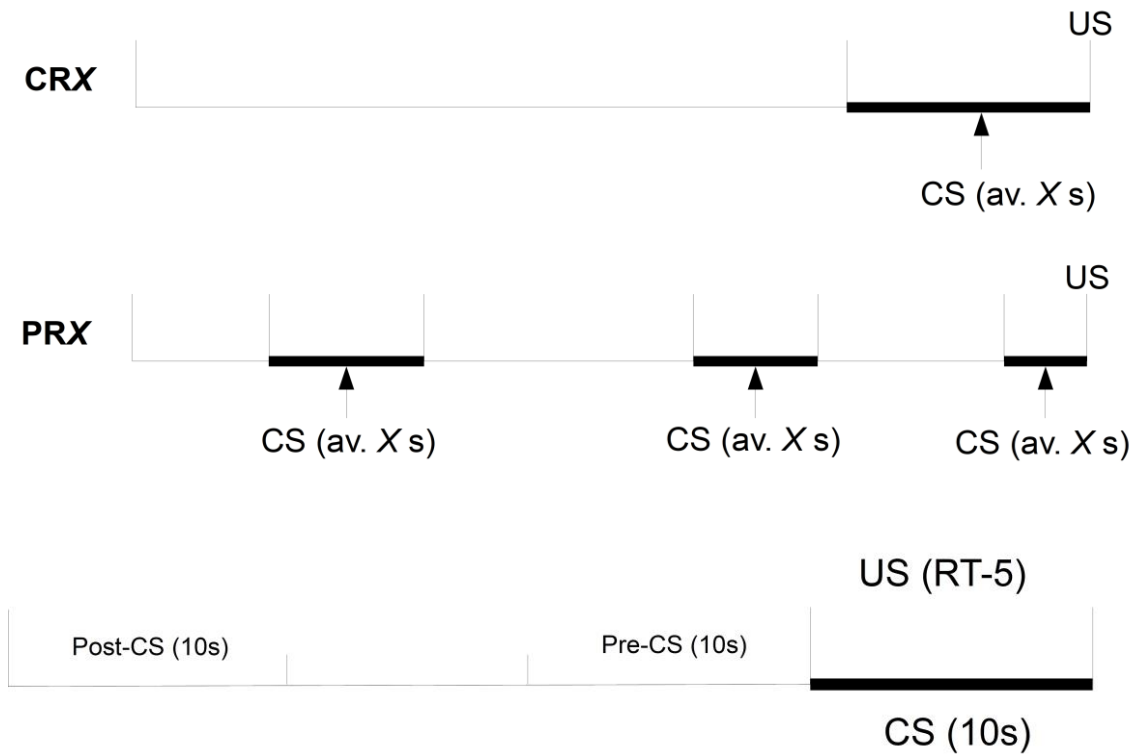

*Figure 3: Task diagrams for Experiment 2 of Harris et al. (2015) (top) and the experiment of Calcagni et al. (2020) (bottom). Here X=10 or 30.*

## 4. Model Selection: RW, DOM, and AOM

Having several theoretical models at hand, we want to see whether one of them explains data better than the others. The first step is to find the best fit of individual data for each model; the second step is to compare these best fits.[5]

In this section, the RW model, the DOM, and the AOM are treated democratically as three different fitting curves for discrete data. This procedure is fair because, on one hand, one makes the same assumptions for all models, i.e., that they can be used as continuous fitting curves; and, on the other hand, we already showed that the approximation of the sequence of points $v_n$ as a curve $v(t)$ is valid for the large number of trials run in our experiments.

Once the three fits are performed for each subject, they are compared using a model selection procedure. The comparison is made by calculating, for each fit, the Bayesian information criterion (BIC) and the Akaike information criterion (AIC), two quantities that depend on the number of data $N$, the error variance $\sigma_e^2$ (sum of squared residuals) and the number of free parameters $p$:

$$\text{BIC} = N \ln \sigma_e^2 + p \ln N, \qquad \text{AIC} = N \ln \sigma_e^2 + 2p. \qquad (13)$$

---

[5] A best fit is declared significant when the *p* value associated to each of the best-fit parameters is below a certain threshold (for instance, 0.05 or lower). When *p* is above this threshold, it means that there is a non-negligible chance that the search routine could not determine a best fit and that assigned the output value for that parameter randomly. Consistently, the standard deviation associated with that parameter is of the same order of the central value of the parameter itself.

See Calcagni et al. (2020) for more details and references. The better the fit, the smaller the IC and the greater the number of free parameters, the greater the IC. The difference between BIC and AIC is in the way the number of parameters is penalized. For any given subject, the model with smaller IC is the most favored, independently of its theoretical justification.

In practice, the RW model has two free parameters (the asymptote of learning $\lambda$ and the product of the US and CS salience $\alpha\beta$, here considered as one parameter), the DOM has four ($\lambda$, $\alpha\beta$, the oscillation frequency $\mu$, and the amplitude $A$), and the AOM has five ($\lambda$, $\alpha\beta$, $\mu$, and two amplitudes $A$ and $B$). A priori, a fitting curve with more free parameters will fit data better, but it will also be penalized more severely in the AIC and BIC. In other words, the AIC and BIC take into account and partially compensate the advantage of the AOM over the DOM, and both over the RW model, for having more parameters available.

Preliminarily, we checked some convergence issues of the AOM. We performed a nonlinear best-fit analysis[6] for all subjects with the parameter $\mu$ constrained to be positive, and the parameters $A$ and $B$ constrained in several ways ($A \neq 0$, $B = 0$; $A = 0$, $B \neq 0$; $A \neq 0$, $B \neq 0$). However, the problem is that the oscillations of the AOM are not damped in time, and this forces the nonlinear fit algorithm to solutions that are unsatisfactory in all cases regardless of the priors on $A$ and $B$, in the sense that they are characterized by small-amplitude (comparable with error bars), high-frequency oscillations which have little to do with the behavior of the subjects. In general, the IC associated with these fits is much larger than RW or the DOM. There are some cases where the AIC is smaller than the other fits, but they clearly show the above-mentioned artifacts. In fact, one cannot use the AOM as a model of short-range fluctuations either, since a spectral analysis can prove that these fluctuations are random (Calcagni et al., 2020), while the AOM is completely deterministic.

These results are just artifacts of the prior $\mu > 0$ given above and they can be fixed by noting that all oscillatory best fits in the previous notes have very low-frequency oscillations with $\mu < 0.1$. Therefore, we implemented the prior $0 < \mu < 0.1$, plus the priors $A \neq 0$, $B \neq 0$. Fits where $\mu$ was close to 0.1 were double-checked with a longer prior range. Also, in cases where one of the amplitudes of the fit was zero within the fit error, as a double check, we redid the analysis by setting that amplitude to zero as a prior, in order to reduce the penalty from the number of free parameters. Fits with $\alpha\beta$ very large are avoided by imposing the prior $0 < \alpha\beta < 1$, while best fits with $\alpha\beta$ approximately zero within the error are ruled out as artifacts. Oscillatory fits yielding bigger IC and those that fail to produce a nontrivial model (i.e., if $\mu$ and/or $A$ vanish) are discarded too. More precisely, those oscillatory fits with $A$ as a free parameter which are trivial because *A* vanishes are discarded, and the comparison is made only between RW and the DOM with *A*=0 set *a priori*. While subjects where the oscillatory fit is trivial because $\mu$ vanishes are adjudicated to RW.

---

[6] We recall that nonlinear regression is the fit of data with a nonlinear function depending on some free parameters. The problem is reduced to a linear one by suitable transformations or approximations, depending on the functional form of the fitting curve. To find the best fit, one uses the least squares method. For each choice of free parameters, one calculates the sum of square residuals (the vertical distance between each data point and the curve). The set of parameters yielding the lowest sum of square residuals corresponds, by definition, to the best fit.

Before entering into the IC analysis, we would like to stress that the DOM is an extension of RW, rather than a competitor. However, fitting only the DOM and focusing on the magnitude of the parameter $\mu$ instead of comparing the DOM and RW with ICs would be insufficient to establish the presence of oscillations as an empirical phenomenon. Let us see why using the data of Harris et al. (2015). In Fig. 4, we plot the value of $\mu$ for the best fit of the DOM to all subjects.

As one can see in this figure, when the error bar of $\mu$ is large enough to cross the zero value, the DOM is never favored. Conversely, when $\mu$ is nonzero within the error estimate, it should still be compared with RW and with the AOM, because it may be that the dispersion of the data with respect to the fitting curve of the DOM be greater than for other models. In these cases, the ultimate verdict lies in the value of the ICs. This shows that the IC analysis is necessary and cannot be replaced by an analysis of just the presence-vs-absence of oscillations via an estimate of the values of $\mu$. A significant nonzero $\mu$ does not always guarantee the actual presence of oscillations in data (cases CR10-9, CR10-12, CR30-2, CR30-15, PR10-2, PR10-10), nor failure of the DOM implies the absence of oscillations (cases CR10-2, CR10-4, CR10-13, CR30-9, PR10-3, PR30-9). A merit of presenting data as in Fig. 4 is to show that oscillations are present in a large number of subjects.

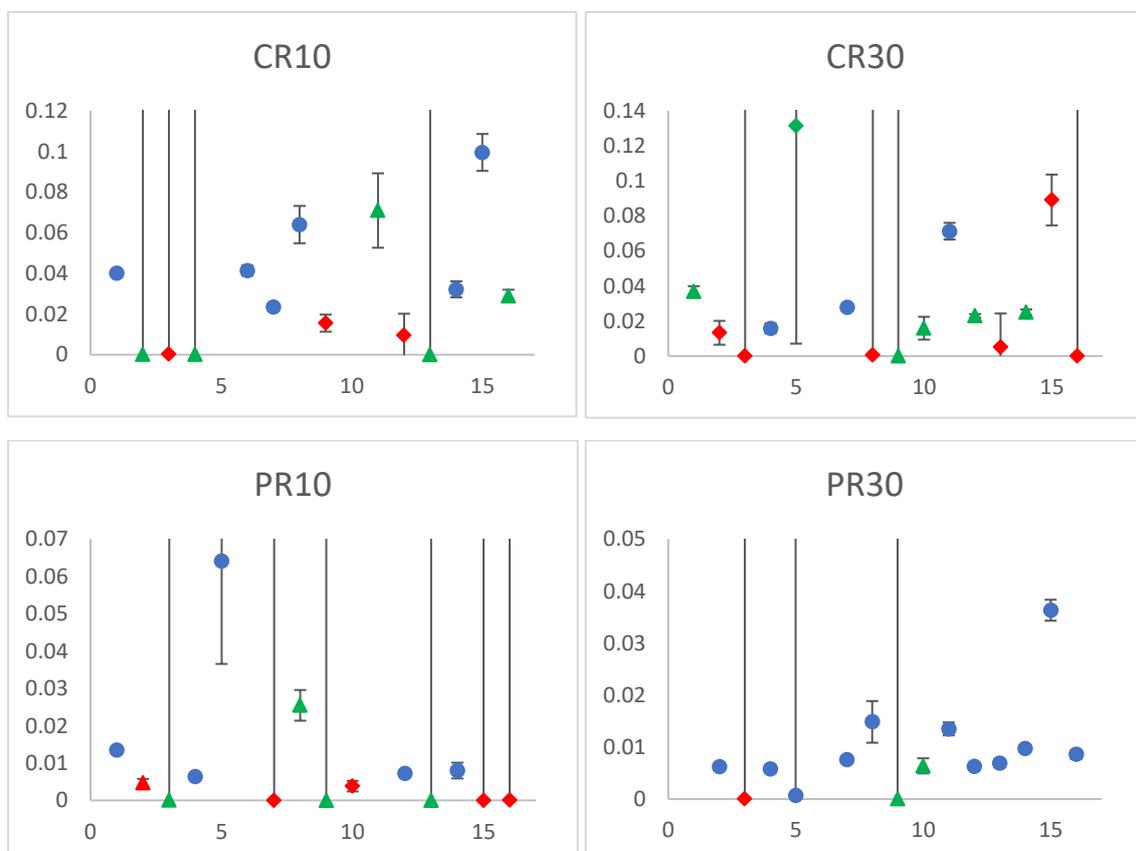

*Figure 4: Values of $\mu$ (*vertical axis*) of the DOM best fit for subjects 1-16 (*horizontal axis*) of the four groups of the experiment of Harris et al. (2015). Blue dots, green*

*triangles and red diamonds correspond to fits favoring, respectively, the DOM, the AOM and RW. The points corresponding to subjects CR10-10, PR30-1 and PR30-6 are off scale and are adjudicated to, respectively, the AOM, the AOM and the DOM.*

### 4.1 Analysis of Harris et al. (2015) Data

The AIC for the DOM and AOM best fits of the data of the subjects of this experiment is shown in Fig. 5. The tables B1-B4 (Appendix B) include more details of the Bayesian analysis, including the BIC.

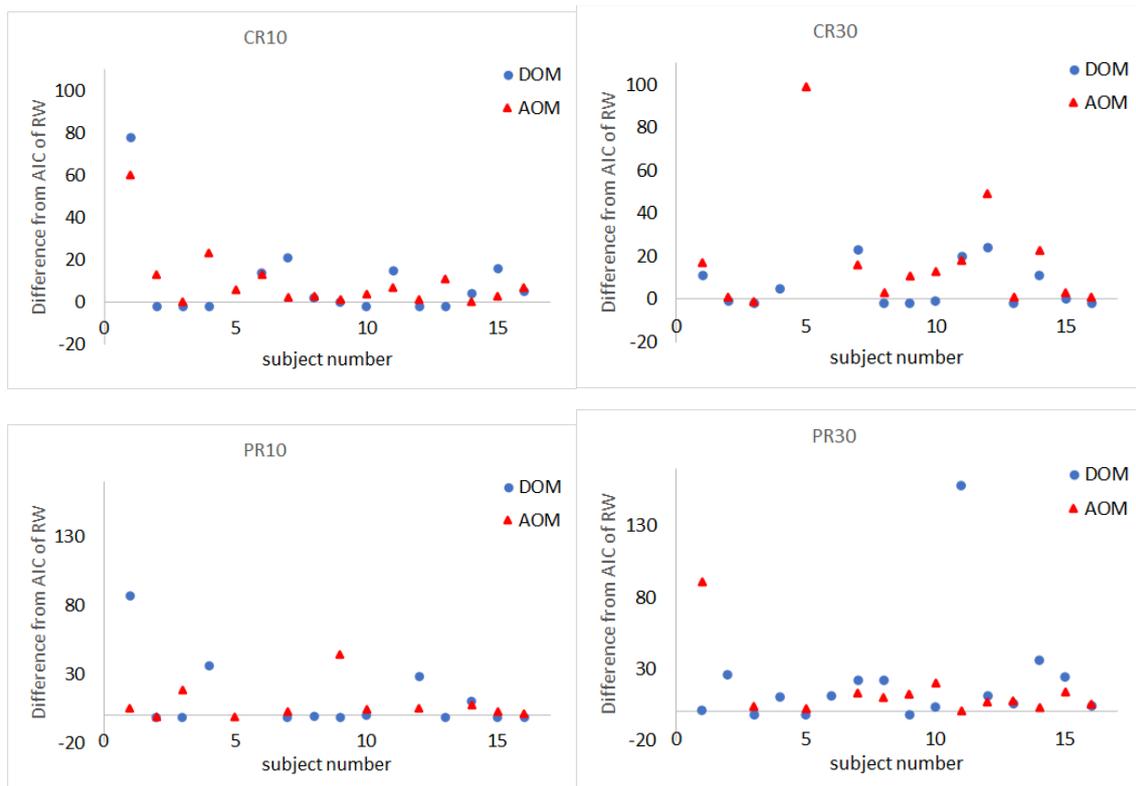

*Figure 5: The difference between the AIC of the RW model and the DOM (blue circles) or the AOM (red triangles) for the 16 subjects of Harris et al. (2015). Positive values indicate that the DOM/AOM is a better fit than the RW model, while for negative values the RW model is a better fit. Comparing the DOM and the AOM, the fit with better evidence is the one with larger value. Cases where one could not fit with the DOM or AOM have no corresponding point.*

From these figures and tables, we observe that the best fit of the majority of subjects is either the DOM or the AOM (with a predominance of the DOM, although this is not clear from Fig. 5 or its BIC counterpart and one should use the combined information of both, as explained in Appendix B; see Table 1), not the RW model. When the RW model wins it does so with weak to positive evidence in CR groups and weak to strong evidence in PR groups. This increase in evidence in favor (but not in the number of favored cases) is probably due to the larger number of data points (three times as many in PR groups with respect to CR groups). On the other hand, the great majority of oscillatory winners have very strong evidence in favor.[7] Thus, when subjects display oscillations in their learning curve the effect is usually strong, not just tiny ripples around the monotonic RW curve. In other words, oscillatory models typically win with strong or very strong evidence because they fit data remarkably better than RW, which means that the oscillatory pattern in data deviates considerably from the monotonic RW curve.

A group view of these results is summarized in Table 1.

|  | CR10 | CR30 | PR10 | PR30 | Total |
|---|---|---|---|---|---|
| **RW** | 19 % | 37 % | 31 % | 13 % | 25 % |
| **DOM** | 44 % | 19 % | 31 % | 69 % | 41 % |
| **AOM** | 38 % | 38 % | 25 % | 19 % | 29 % |
| **No fit** |  | 6 % | 13 % |  | 5 % |

*Table 1: Percentage of subjects following the RW model, the DOM, or the AOM, using the combined information of the BIC and the AIC (see Appendix B for details). The learning curve of subjects CR30-6, PR10-6, and PR10-11 did not follow any of the models.*

It is important to note that RW is a subcase of the DOM and of the AOM. This means that, whenever RW is selected as the most favored fit, then the DOM and the AOM with frequency approximately zero may also be good fits, while if one of the oscillatory models wins, then the frequency cannot be set to zero and RW is not a good fit. Given that the DOM and the AOM have RW as a reduced form, they are guaranteed to be at least as good as RW if we do not penalize them for having more free parameters. However, indeed they are penalized by the BIC and the AIC and one cannot invoke chance to trivially explain why they fare better than RW for some subjects. Therefore, since the goal is to fit as many subjects as possible with the *same* model, Table 1 leads us to conclude that the DOM provides a more powerful explanation of the data than RW.

Table 1 is helpful also to look for inter-group differences. In CR groups (continuous reinforcement), the number of subjects following the RW model and the AOM increases when extending the trial duration, while the number of those following the DOM decreases. In PR groups (partial reinforcement), the opposite occurs: the number of subjects following the RW model and the AOM decreases when extending the trial duration, while the number of those following the DOM increases. This puzzling pattern received an explanation in Calcagni et al. (2020), without including the AOM. To summarize it here, it seems that *longer trials stabilize the behavior, but a partial reinforcement schedule destabilizes it*. Here, stability means fewer erratic patterns and more monotonic (RW) learning, but it is not accompanied by smaller oscillations in oscillatory patterns.

This interpretation implies an ordering of the groups in terms of increasing stability:

$$PR10 \rightarrow PR30 \rightarrow CR10 \rightarrow CR30 \qquad (14)$$

---

[7] As recalled in Appendix B, very strong evidence of a model over another is defined as a difference in the IC equal or greater than 10 (Jeffreys, 1961; Kass and Raftery, 1995). Most dots/triangles above the horizontal axis in Fig. 5 have a vertical coordinate greater than 10. Dots/triangles below the horizontal axis correspond to subjects for which the RW model wins, and these points are always very close to the axis: This is the graphical illustration of the fact that, when RW wins, it never does it with very strong evidence.

Including also the AOM, we can check if data reflect this feature by constructing three orderings from Table 1: One where the number of subjects following RW increases and those following the DOM decreases, PR30 → CR10 → PR10 → CR30, another where the number of subjects following the AOM increase, PR30 → PR10 → CR10 → CR30, and a third one where the number of subjects with erratic behavior increase: PR30 ~ CR10 → CR30 → PR10. Except for the position of PR10 in the sequence, (14) is essentially confirmed. A fourth way is to sum all the estimated error variance with respect to the best-fit curve of each subject. This procedure confirms Eq. (14) except for the position of PR30. Therefore, the hypothesis that there is an order of increased behavioral stability determined by the trial length and the reinforcement schedule, such that the longer the trial and the more continuous the reinforcement, the more stable and monotonic the learning curve, is only partially confirmed.

Finally, in Figs. 1 and 6-9 we report the best fits for all subjects.

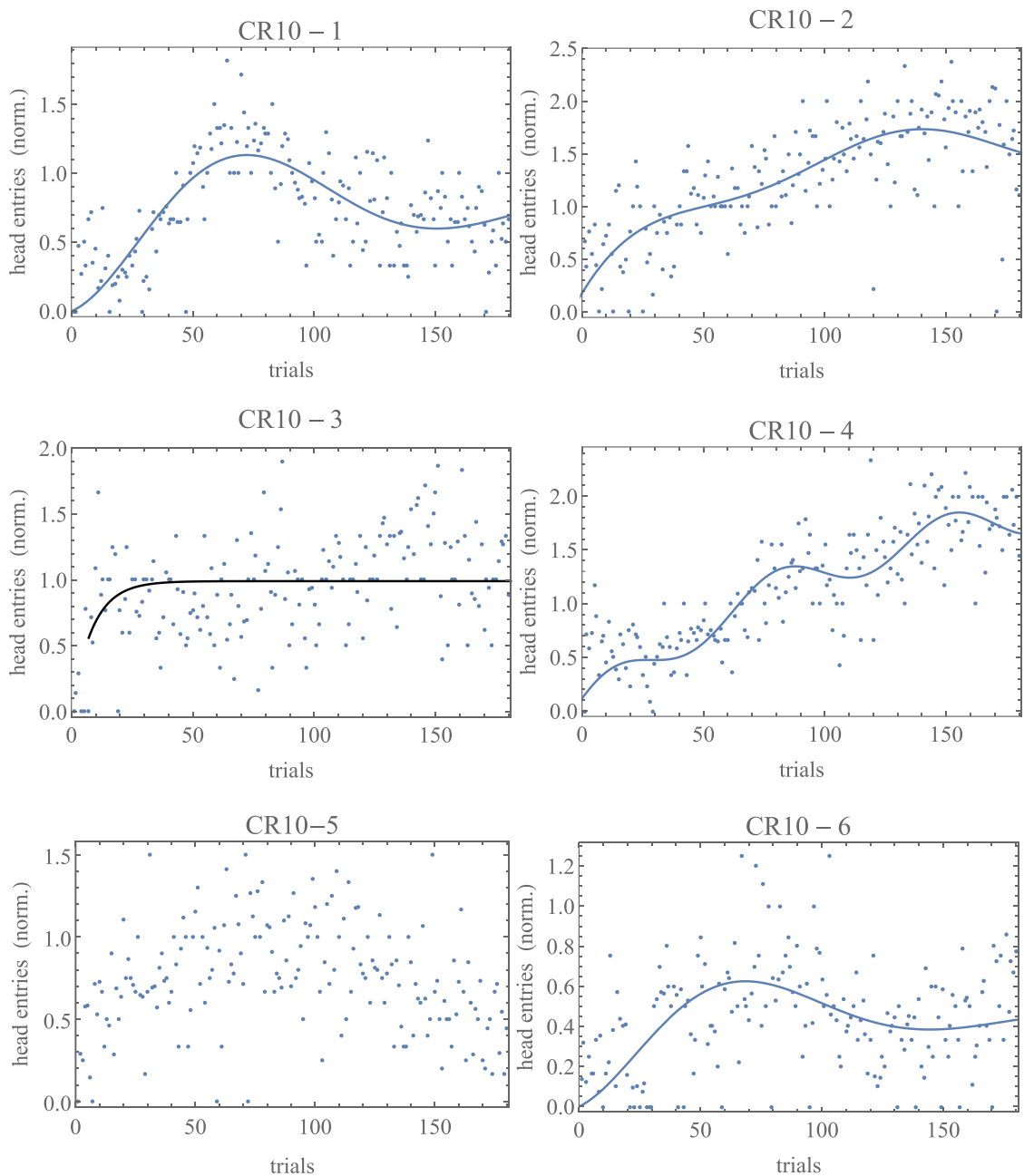

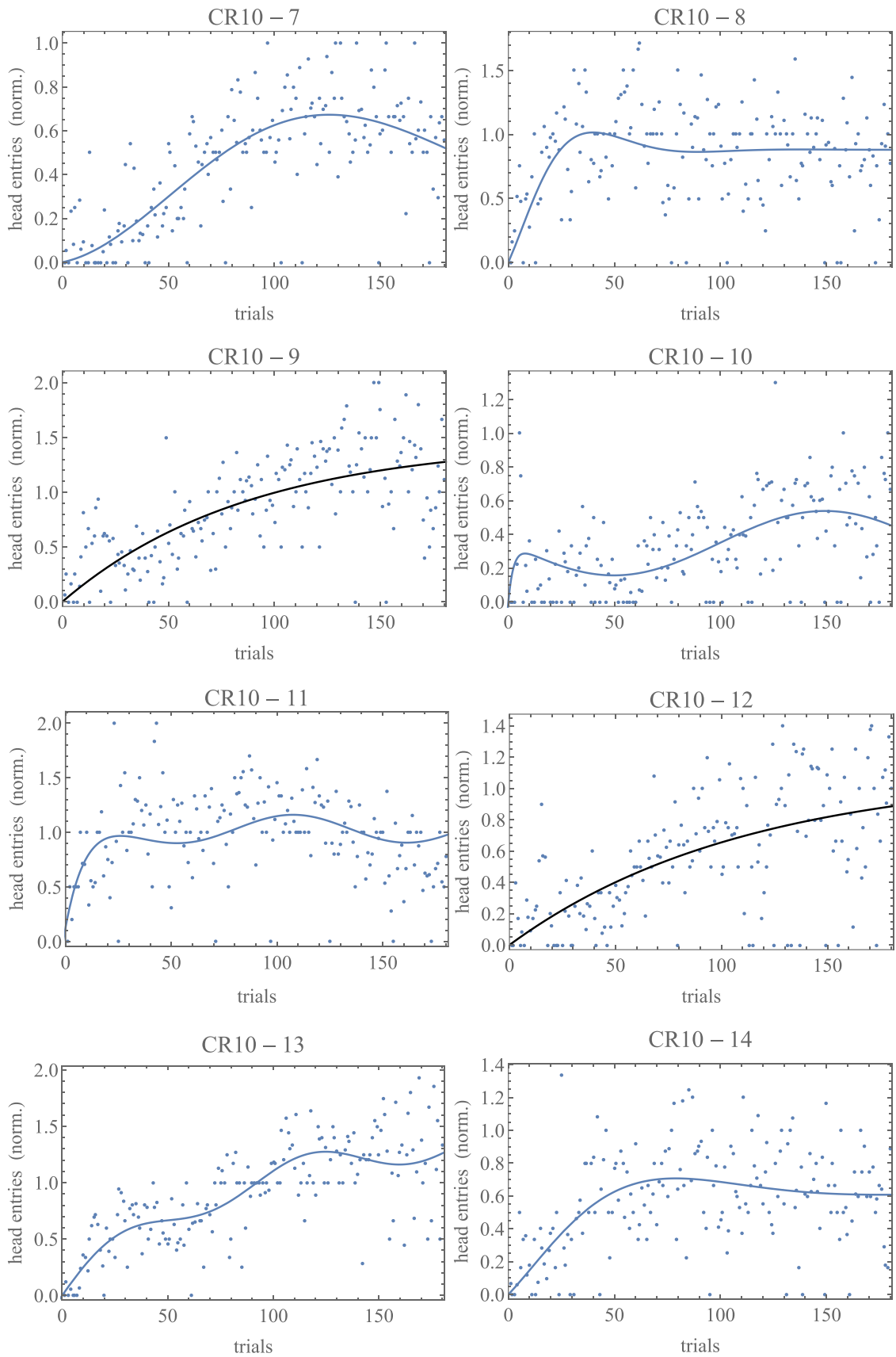

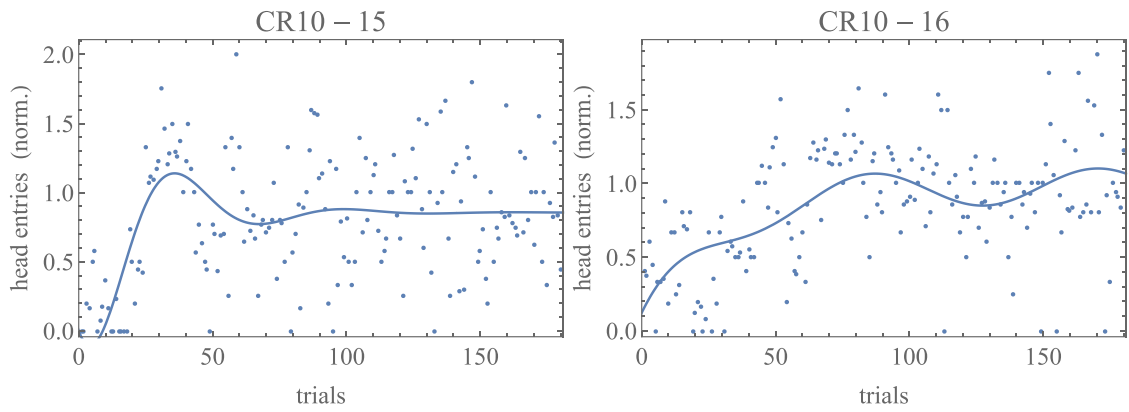

*Figure 6: Best-fit plots for the subjects of group CR10. RW: black line; DOM/AOM: blue line. Horizontal axis: trials (1 to 180). Vertical axis: normalized response (with respect to the asymptote of learning; see Calcagni et al. (2020) for details). Case 5 has no fitting curve because the best fit is not significant (see Appendix B).*

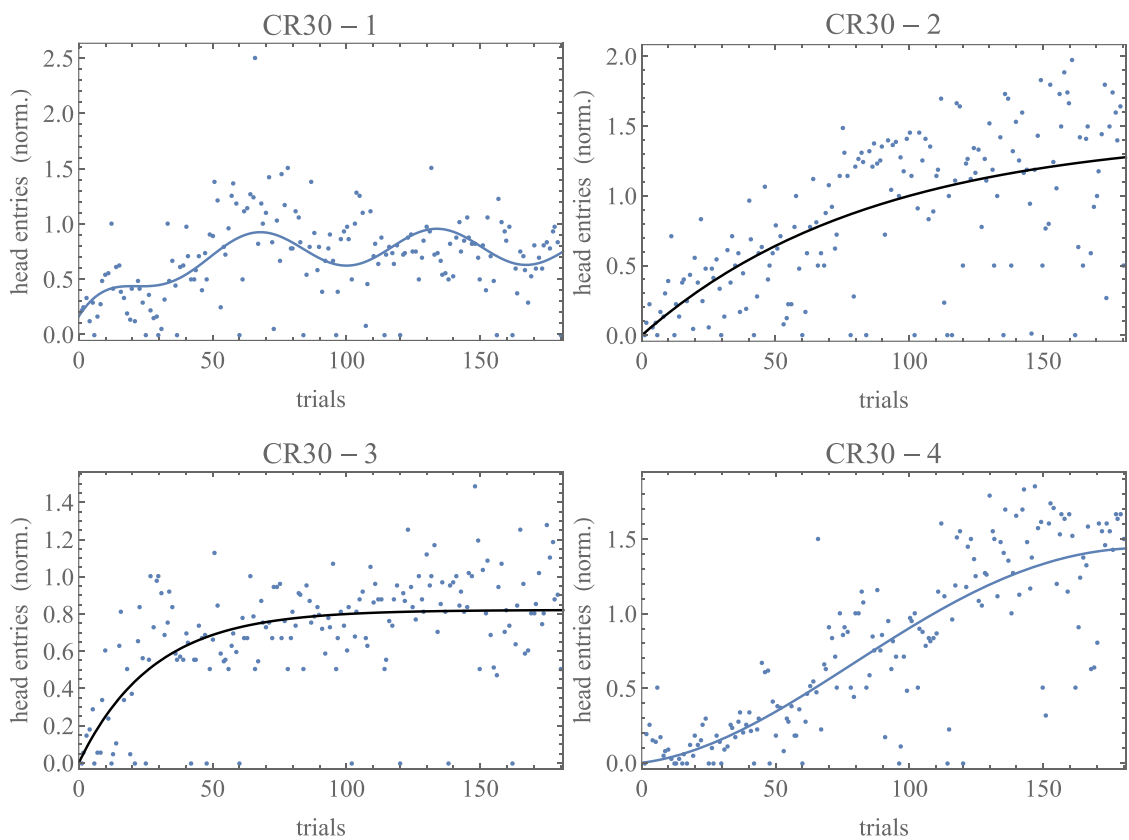

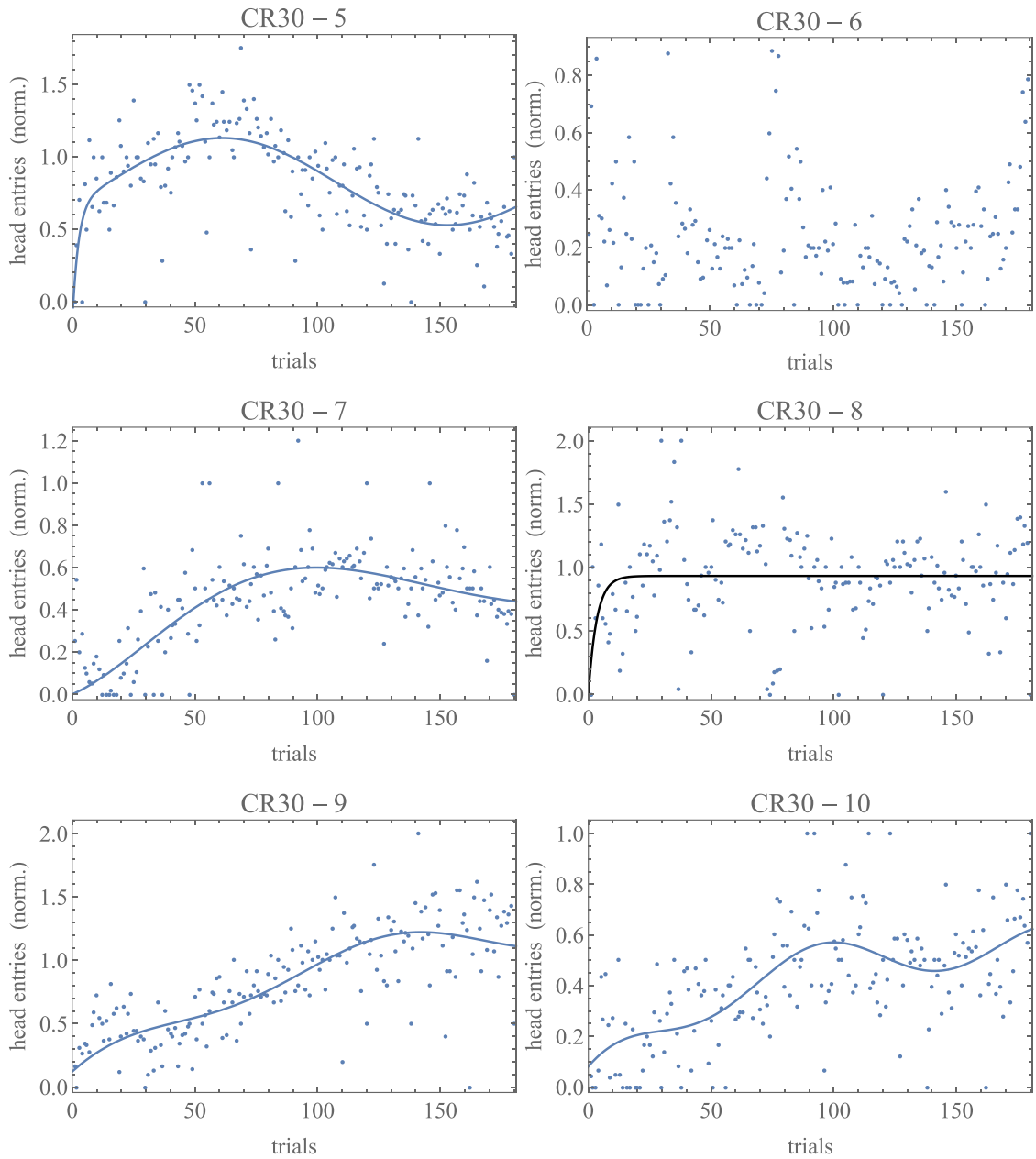

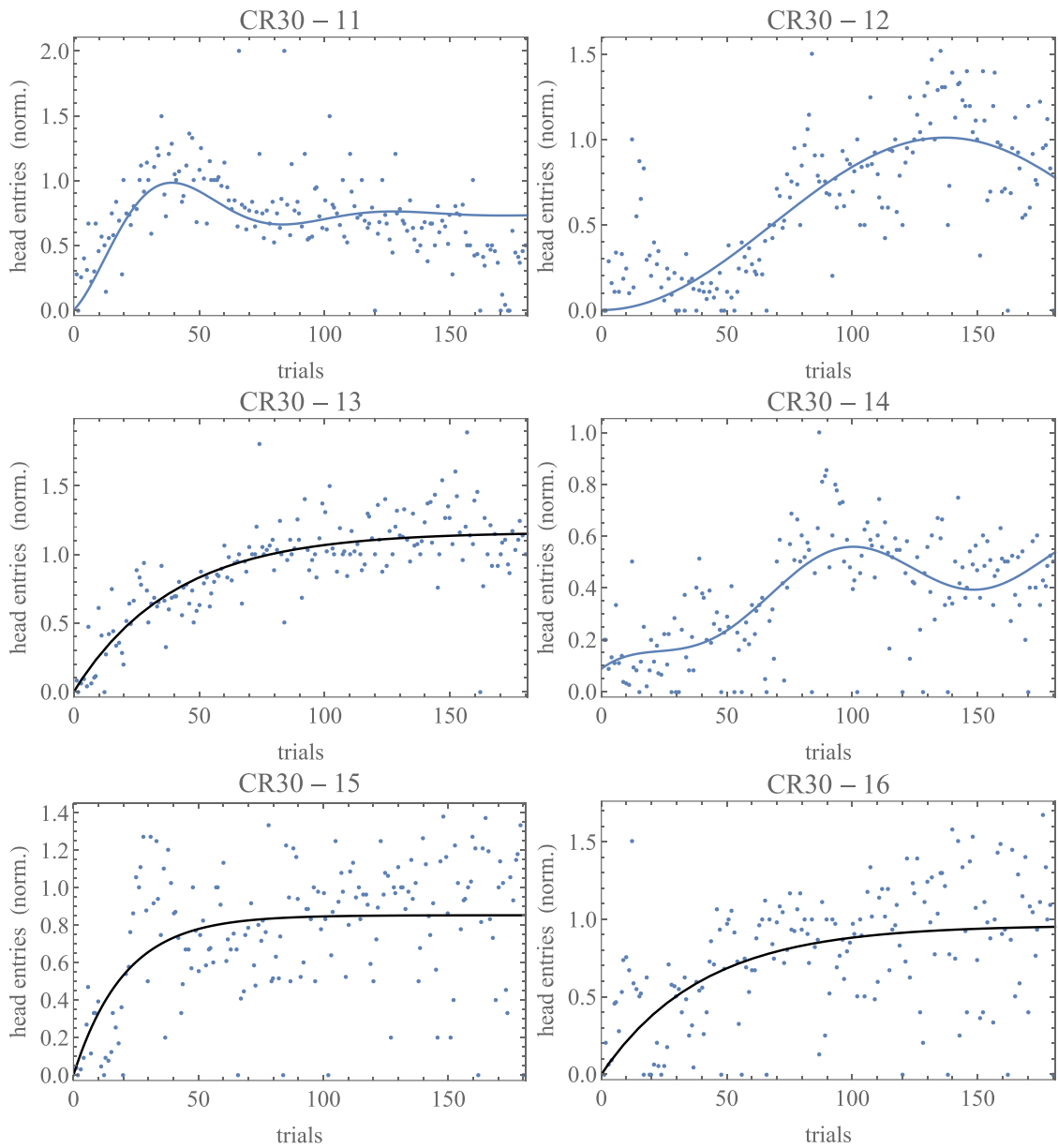

*Figure 7: Best-fit plots for the subjects of group CR30. RW: black line; DOM/AOM: blue line. Horizontal axis: trials (1 to 180). Vertical axis: normalized response. Subject 6 has no fit.*

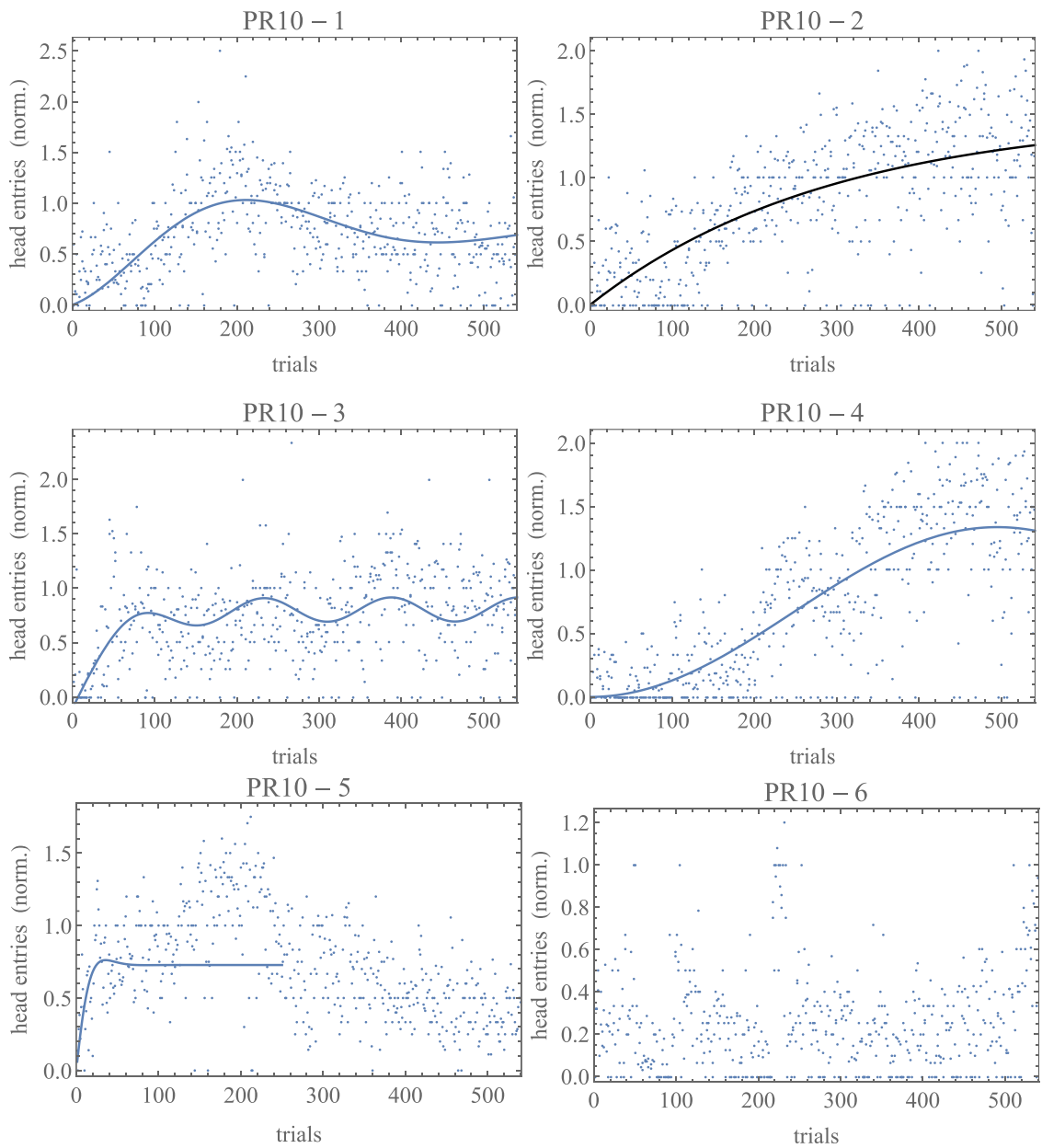

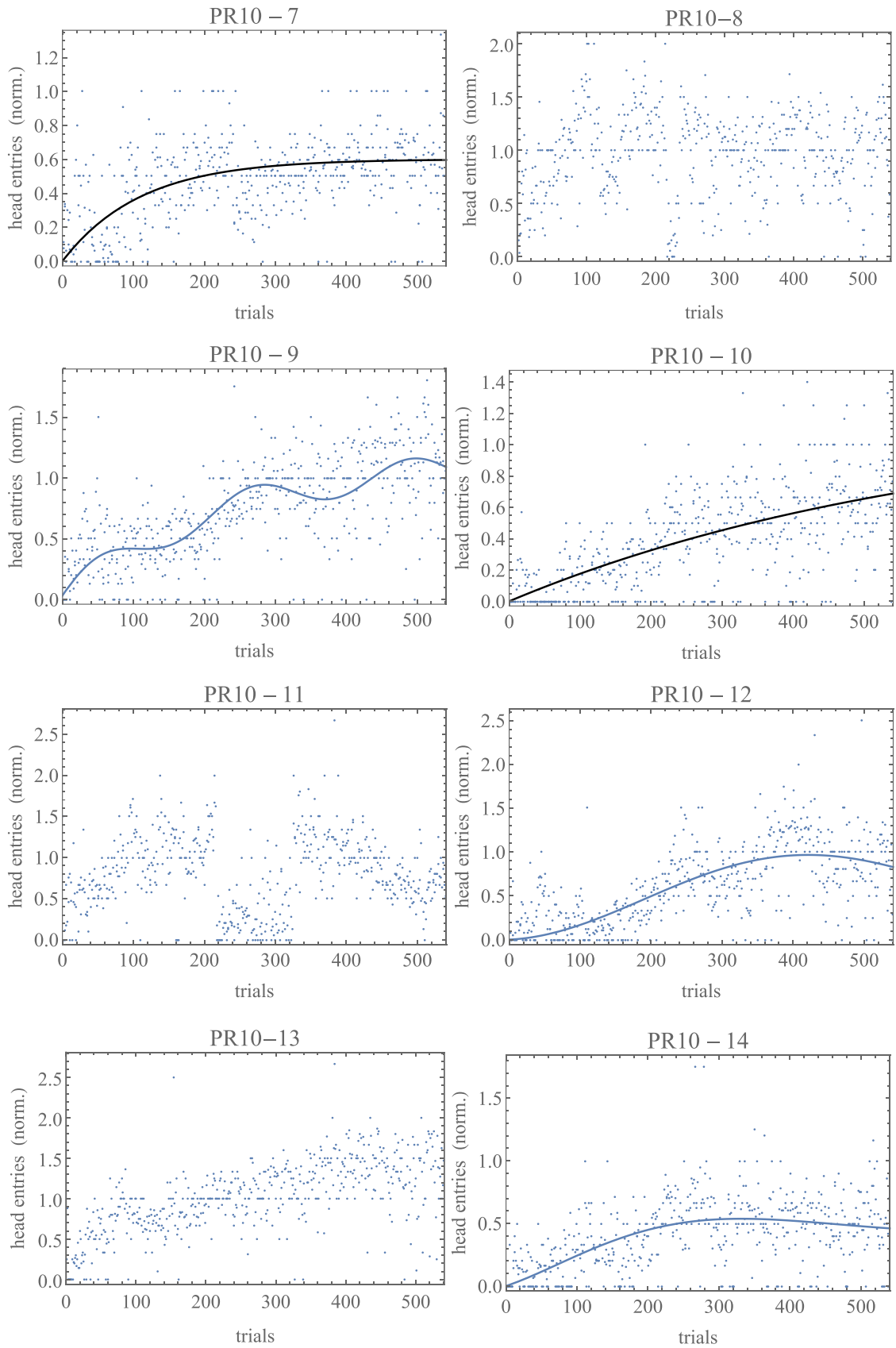

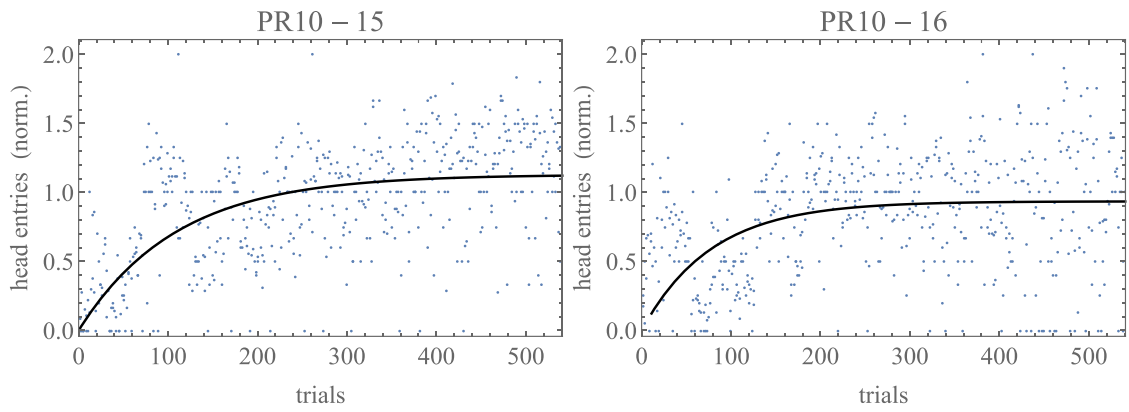

Figure 8: *Best-fit plots for the subjects of group PR10. RW: black line; DOM/AOM: blue line. Horizontal axis: trials (1 to 540). Vertical axis: normalized response. Cases 6, 8, 11 and 13 have no fitting curve because the best fit is not significant (see Appendix B).*

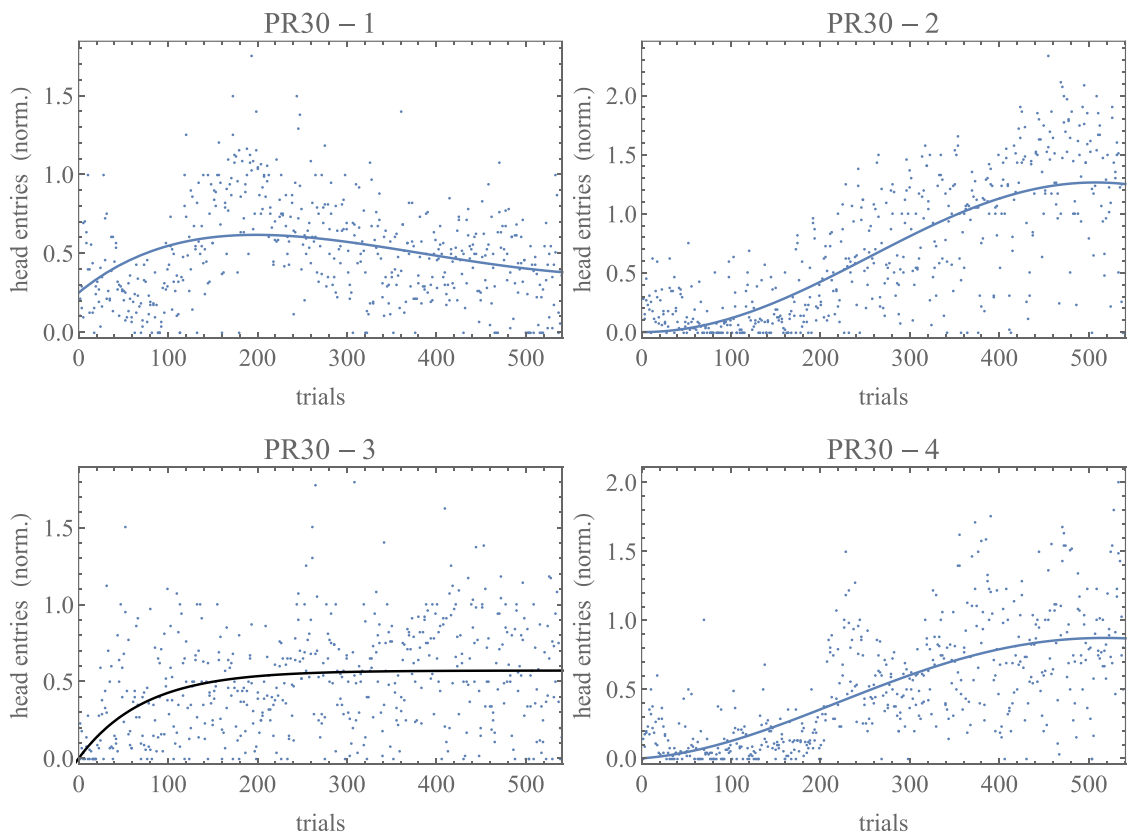

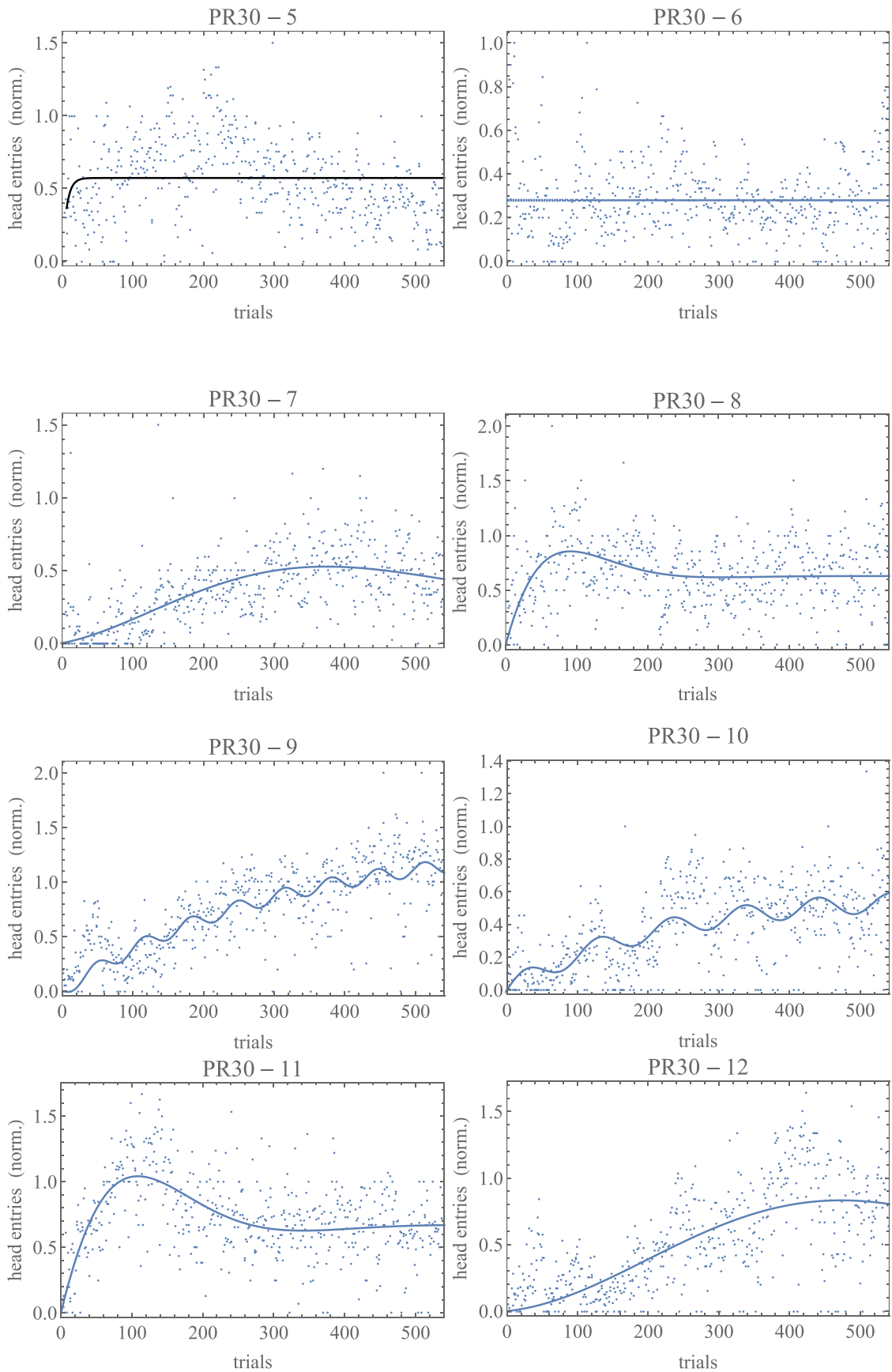

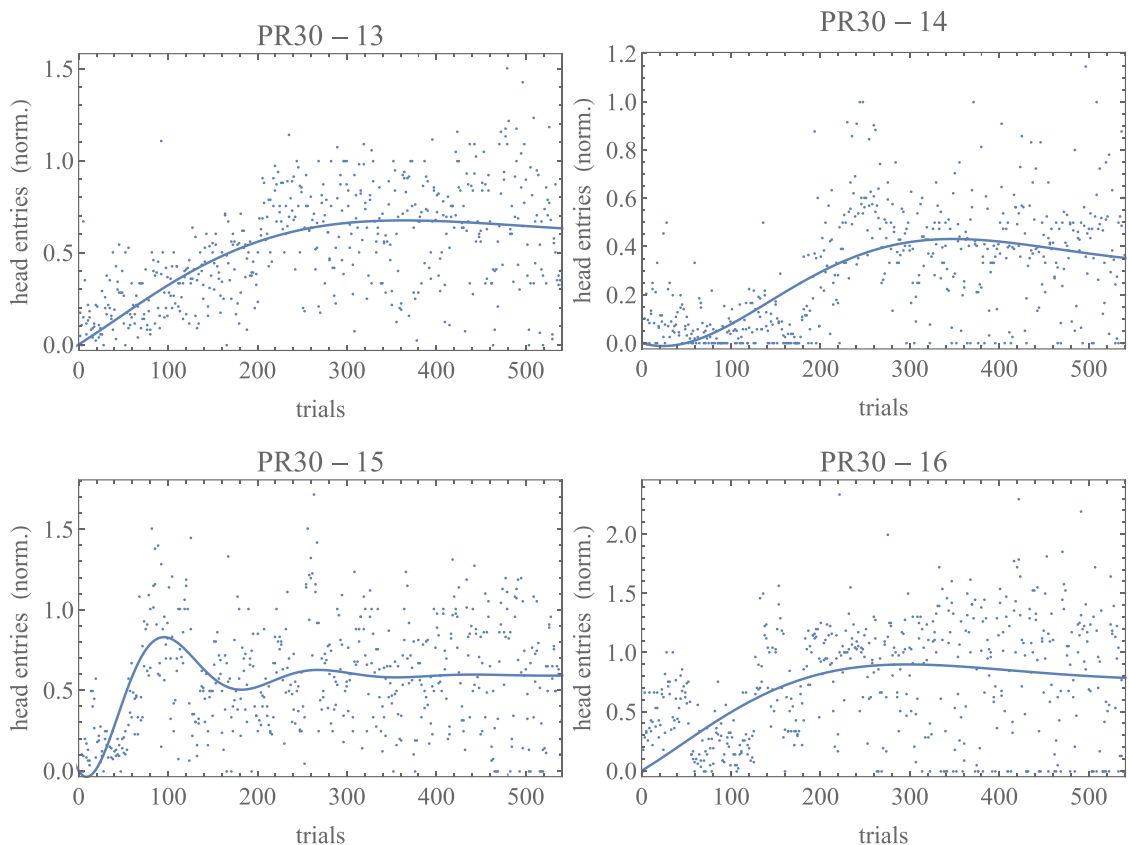

*Figure 9: Best-fit plots for the subjects of group PR30. RW: black line; DOM/AOM: blue line. Horizontal axis: trials (1 to 540). Vertical axis: normalized response.*

As a final remark, we treated the data of Harris et al. (2015) as if all sessions were concatenated. One may wonder whether response decay between sessions, within-subject or other between-session differences could cause oscillatory patterns such as those ascribed to the DOM. In Calcagni et al. (2020), we argued that the underlying mechanism of periodic "smooth" oscillations is very different from sources of short-scale random fluctuations. We quantitatively illustrated the point exploring random-fluctuation models together with the DOM. Let us discuss the issue here with a different wording. It is well known that within-sessions performance is not fully stable, especially when sessions are long enough, and that performance may be affected by inter-session intervals. However, in the present case the scale of the effect is different from the one of oscillations. For example, in Fig. 6, subject CR10-4, the oscillations have a semiperiod (length from a minimum to a maximum) of about 50 trials, to be compared with the session duration of 6 trials (for a total of 30 sessions = 180 trials). In Fig. 7 CR30-1, the semiperiod is of about 25 trials, again to be compared with 6-trial sessions. In Fig. 8 PR10-3 and PR10-8, the semiperiod is of about 100 trials and 50 trials respectively, to be compared with sessions of 18 trials. So, in all cases the scale of the oscillations is 3 to 9 times larger than the inter-session interval. This means that any performance-decay effect would take place at scales much smaller than those of oscillations and, on top of that, it would produce a sawtooth-like or even a random noise-like modulation of an otherwise monotonic learning curve. As a matter of fact, this might as well be one of the underlying mechanisms explaining data dispersion in empirical learning curves. Within-subject differences would produce pretty much the same pattern: random short-scale fluctuations rather than coherent long-range oscillations. Furthermore, we would expect inter-session effects to produce about the same pattern in all subjects, at least qualitatively. However, many subjects display oscillations with a much longer period (thus increasing the causal gap between long- and short-scale effects), while others show no significant oscillation at all (thus weakening the hypothesis that inter-session effects could invariably produce sizable periodic oscillations).

### 4.2 Analysis of Calcagni et al. (2020) Data

Here we do not repeat the analysis done in Calcagni et al. (2020) but we expand it to include the AOM best fits.[8] The results are in Fig. 10 and Tabs. B5 and B6. On the one hand, the AOM never beats the DOM when the latter beats RW. This reinforces the view that the DOM is typically a better fit than the AOM. On the other hand, in most cases, the RW model still beats the oscillatory models, probably because of the heavy penalty on the latter (4 to 5 parameters versus 2).

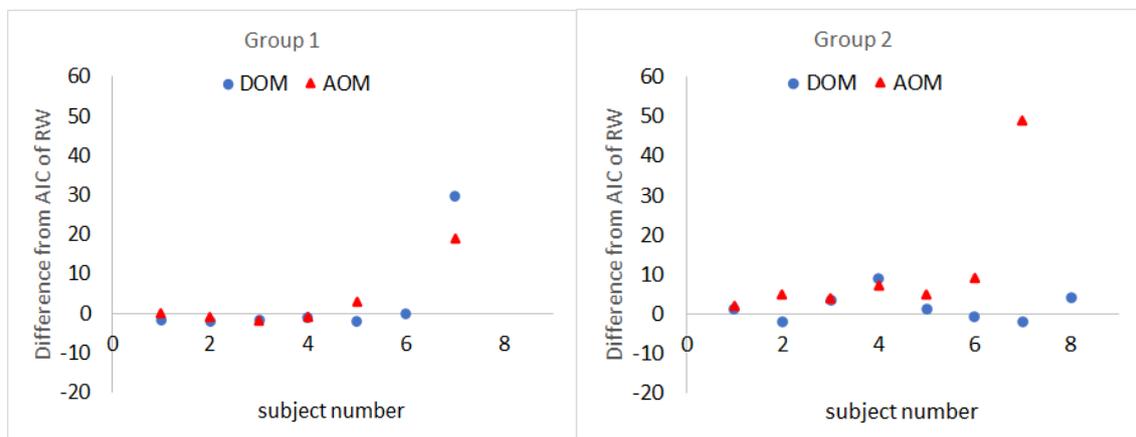

*Figure 10: The difference between the AIC of the RW model and the DOM (blue circles) or the AOM (red triangles) for the 15 experimental subjects of Calcagni et al. (2020). Positive values indicate that the DOM/AOM is a better fit than the RW model, while for negative values the RW model is a better fit. Comparing the DOM and the AOM, the fit with better evidence is the one with larger value. Cases where one could not fit with the DOM or AOM have no corresponding point. The point of subject 2-8 corresponding to the AOM is at a very negative vertical position and is not shown.*

There are four cases (subjects 2-2, 2-5, 2-6, and 2-7) where the AOM beats RW (and the DOM is beaten by RW). These exceptions do not jeopardize the overall picture of a successful DOM, the reason being that we may expect an ad hoc model to fit some data better than RW and the DOM. Nevertheless, it is still remarkable that the DOM and the AOM can fare better than RW despite having at least twice as many parameters.

---

[8] Note that the fits of the oscillatory and RW models of this subsection were made with session data (90 points), while those in section 5.1 were made with trial data (180 or 540 points). Trial data of Calcagni et al. (2019) have larger dispersion than Harris et al. (2015) trial data, while their dispersion is comparable when the former data are binned into sessions.

Figure 11 includes only subjects for whom the best fit is the AOM; the other fits can be found in Calcagni et al. (2020).

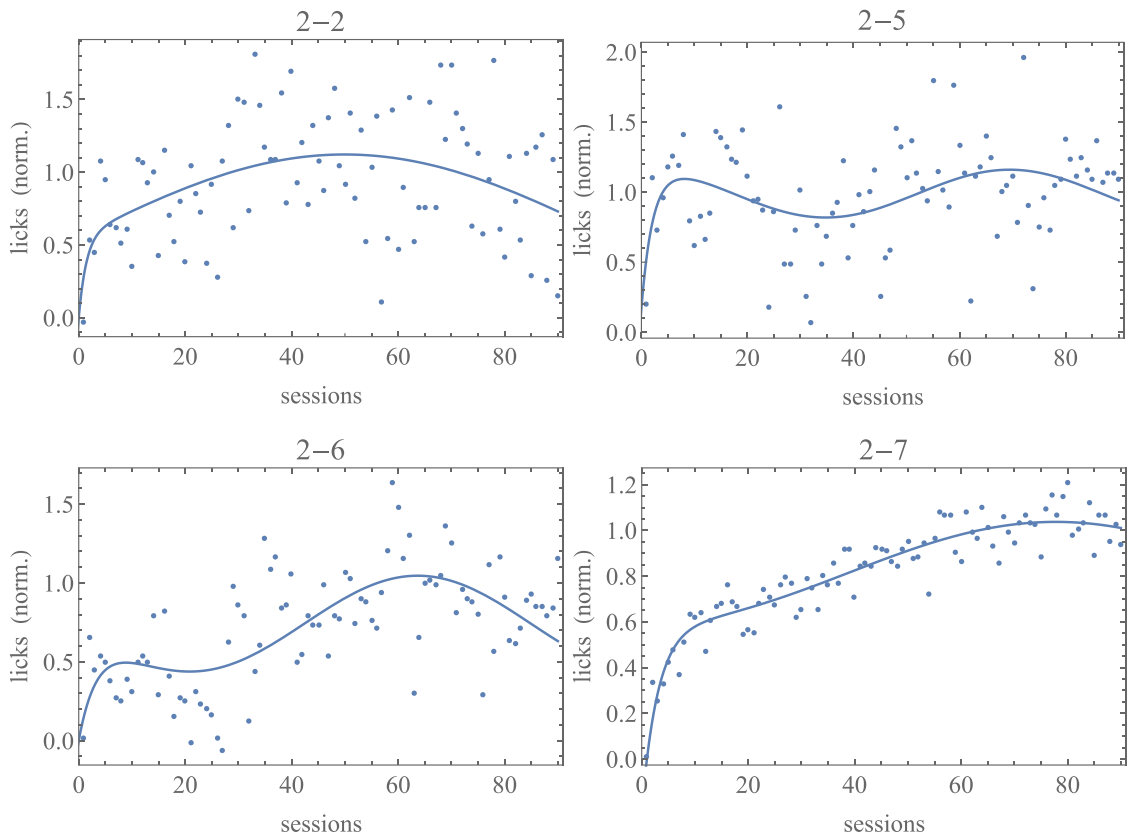

*Figure 11: Subjects (all belonging to Group 2) where the AOM is the most favored model (evidence is against the RW model in all cases. Horizontal axis: sessions (1 to 90). Vertical axis: normalized response (with respect to the asymptote of learning; see Calcagni et al. (2020) for details).*

The percentages of subjects following the RW model, the DOM, or the AOM, are shown in Table 2. As one can see, in Group 1 the AOM is absent, while in Group 2 it explains half of the data, better than the other two models. Overall, the RW model explains slightly less than half of the data, while the AOM explains as many subjects as the DOM, about three quarters of the total number of subjects (remember that the RW model is a subcase of both the DOM and the AOM, so that one should sum the RW percentage to that of the DOM or AOM).

|  | Group 1 | Group 2 | Total |
|---|---|---|---|
| **RW** | 86 % | 12 % | 47 % |
| **DOM** | 14 % | 38 % | 27 % |
| **AOM** | 0 % | 50 % | 27 % |

*Table 2: Percentage of subjects following the RW, the DOM, or the AOM.*

The fact that the US in Group 2 was twice as sweet as that in Group 1 (and that was preferred by rats in a pilot discrimination experiment) endorses the theoretical view that the AOM is a motivational model rather than an associative one. Still, it remains difficult to understand where such an oscillatory motivational pattern might come from. If oscillations were due to motivational factors (performance versus learning), then one would like to keep track of these factors. The first thing one can check is hunger. In this experiment, all subjects were not food-deprived and were kept at 100% of their theoretical weight (calculated by an ideal growth curve). Food was delivered in the cage at least half an hour after the last session of the day for the rat. Hunger does not appear to be a relevant factor on the oscillations that took place in the range of several experimental sessions and not locally determined by individual sessions.

It is also true that the saccharin dispensed in the experimental sessions was more appreciated than the regular food in the cage. Therefore, some additional incentive properties may be attributed to the reinforcer in our experiment. If this were the case, the determinants of the level of oscillations might be critical manipulations related to hunger and to incentive properties of the reinforcer. Additionally, oscillations should be greater at the start and end of long sessions given fatigue effects. However, these considerations do not apply to our experiment. We did not find any correlation between the session time, response level, and cage food delivery. If there existed a motivational factor, then it would have been something different from food. Moreover, in this experiment, all sessions had the same duration and there was no correlation between the number of sessions per day (either 1 or 2) and the response level. Notice also that any response fluctuation at the beginning or end of a session would be on a far shorter time scale than the long-range oscillations of the DOM and the AOM. The motivational factor we are trying to find must be on scales much larger than that of session quirks. Finally, but somewhat unrelated from the rest, one may ask whether motivational factors are responsible for response fluctuations happening on a few trials or sessions, but response fluctuations (dispersion of data) were constant throughout the whole experiment (Calcagni et al. 2020), just as in the case of the data of Harris et al. (2015). Looking at the tables, one sees that dispersion was not systematically greater for longer trials or sessions.

*4.3 Discussion*

We have seen that both the DOM and the AOM capture aspects of the acquisition data for quite a few rats that are not modeled by the RW learning curve. What they seem to be explaining is fluctuations in levels of responding over a longer time scale than trial by trial. The DOM attributes these fluctuations to an oscillatory behavior of the learning system where the animal goes through a sequence of over- and under-adjusted predictions with respect to the optimal asymptote of learning. As such, the oscillations are largest early on in training and progressively damped. The AOM, on the other hand, implements fluctuations as a not-well-specified motivational effect of a non-associative nature.

To assess which model is favored, we compared them on the cumulative percentages, not on the favor attributed to each model separately. This is because RW is a subcase of the other two. In other words, comparison of percentages is among %RW, %RW+%AOM, and %RW+%DOM. The reason (Calcagni et al., 2020) is that, although the DOM (or the AOM) does not explain why a certain subject has $\mu = 0$ and follows RW while another has $\mu \neq 0$ and displays oscillations, it admits both cases, while RW admits, by definition, only the former.

According to Table 1, the DOM explains more data than the AOM. Together with the difficulty in finding a convincing motivational explanation in either experiment, and with the fact that the presence of long-range oscillations in Harris et al. (2015) data is modulated by the duration of the trial and by the type of reinforcement schedule (partial versus continuous), we may conclude that the majority of the observed response oscillations have an associative origin. The key point is that it is impossible to fit a random phase (as a motivational quirk could produce) with a long-range deterministic pattern. Any motivational factor should be of the same time scale as the oscillations and we can hardly find one in these experiments, where the only change in the animal diet and habits was of order of a few hours.

Convincing evidence that the DOM is better than the AOM is also given by model selection statistics. The AOM has 4 to 5 parameters (depending on whether we set one of the amplitudes to zero), while the DOM has 3 to 4. In general, the AOM beats the DOM always when it has more parameters, while the DOM beats the AOM when it has fewer than or the same number of parameters. This outcome seems to be helped by the penalty from the number of parameters in the AOM, but not much. In many cases, the 4-parameter DOM beats the 4-parameter AOM with very strong evidence, which means that the lion's share in the IC is not the number of parameters but the lower data dispersion with respect to the fit curve. To put it simply, the DOM wins by brute force, not because of its lesser penalty. If one simply looks at the goodness of fit, without taking account the number of parameters, the DOM usually gives better fits than AOM.

So, while the DOM beats the RW model often, adding oscillations to RW in the form of the AOM does give it some advantage, but there are still plenty of cases where the DOM beats the AOM, or is about as good. This is important because the DOM is more constrained. If the best description of responding were really just RW to account for the acquisition part plus an oscillatory process to account for some independent factor (e.g., motivation) that causes fluctuations in performance, then the AOM should beat DOM much more often, even more so in the very long experiment of Calcagni et al. (2020). However, this is not the case. Therefore, the overall evidence favoring the DOM justifies some new description of acquisition that includes long-range oscillations of associative origin.

Of course, one can construct infinitely many mathematical models for learning curves, with or without response variability, but the real novelty of this paper and its predecessor is that we are not doing blind model building but, rather, we are testing the predictions of those models admitted by the least action principle (Calcagni et al., 2020). The DOM is the solution of the simplest, single-cue model following this principle. If we believe in the least action principle as a guidance to create learning models, then one should be able to see long-range oscillations in any situation where the RW model has been applied successfully until now. A reanalysis of data of any such experiment should be easy to carry out under the new paradigm, but this may not be an easy task. The overwhelming majority of studies present group data averaged over a number of subjects, a procedure that invariable washes out oscillations since each subject has a different phase (Calcagni et al., 2020). This is probably the reason why oscillations have been left unnoticed so far. To check the DOM in old data, one should necessarily look at individual trends rather than group-averaged learning curves. We hope that we have built a sufficiently strong case to convince the community to look at their data with a renewed interest.

Another avenue to explore is cue competition. In section 3.2 of the Supplementary Material of Calcagni et al. (2020), we applied the least action principle to multi-cue configurations and, in particular, to the dynamical formulation of the full RW model with many CSs. It would be interesting to introduce the parameter $\mu$ in the potential and consider the multi-cue version of the DOM. The multi-cue RW model is also interesting because it could make oscillations possible without any new parameter. In the simplest case, the context and the context + CS compound form a two-dimensional system that can, in principle, sustain oscillations in the subject's response. Therefore, the question arises about whether multi-cue RW could explain more data (or the same data but better) than the DOM, without any major shift in the paradigm of associative learning. However, we foresee disadvantages in multi-cue RW as a competitor of the DOM, for three main reasons. (1) One should justify the use of multi-cue RW identifying the other cues in the experiment but, if the latter has been designed carefully as a single-cue configuration, there may not be many candidate cues available and, if there are (e.g., the "context"), these are likely to be under a poorer control than the main cue. (2) In its simple version with only two cues, the number of free parameters equals that of the DOM and increases linearly with the number of cues. This means that multi-cue RW would have a much stronger predictive power with respect to the DOM (the greater the number of free parameters, the greater the fit flexibility) but it would also be strongly penalized by the information criteria. (3) Multi-cue RW as a dynamical system can be very difficult to solve exactly and it may require numerical methods to find solutions (i.e., learning curves), while the learning curve of the DOM is an exact solution of the model. That is to say, the DOM is under a much greater analytic and numerical control than multi-cue RW. Although it may not be impossible to treat multi-cue RW numerically, we believe this to be a notable computational disadvantage.

In the last two decades, the RW model has come to prominence in the neuroscientific community thanks to an unsuspected correspondence between its prediction errors for rewards and neural responses in the dopaminergic pathways of the midbrain (Schultz et al., 1997). The question of whether those observations can also be explained by the DOM is complex and we do not have a definite answer. Both the temporal-difference model considered by Schultz et al. (1997) and the DOM are Markovian, that is, the presentation of future CS and US depends only on immediate future cues, not on past ones. However, they differ in the fact that, while this property is an assumption in the temporal-difference model, in our case it is not. We did not make any assumption a priori about the time dependence of the association strength: Eq. (6) is a consequence of the least action principle. Also, subjects following the DOM are "slow learners" in the sense that their response becomes stable later than that of the corresponding RW followers. One could use this feature to check the DOM with neurological data, which would help one to understand under what internal and/or external conditions subjects show response oscillations. A quantitative comparison goes beyond the scope of this publication, but it may deserve future attention.

***Acknowledgments.*** We thank our collaborators in Calcagni et al. (2020) and Harris et al. (2015) for participating in collecting the data we analyzed here, and Stefano Ghirlanda for insightful comments on a previous version of the manuscript. R.P. was supported by grant PSI2016-80082-P from Ministerio de Economía y Competitividad, Secretaría de Estado de Investigación, Desarrollo e Innovación, Spanish Government (R.P. Principal Investigator).

## Appendix A – The AOM is not associative

Equation (12) is the most general solution of the equation of motion

$$\left(\frac{d^2}{dt^2} + \mu^2\right)[\dot{v} + \alpha\beta(v - \lambda)] = 0 \,. \tag{15}$$

Since the AOM stems from a third-order equation of motion, it needs more initial conditions (four in total) than RW and than the DOM to specify the solution (12). This is the reason why (12) has six free parameters, the DOM has five, and the RW model only three.

Comparing with (3), it is clear that the RW model is a special case of the AOM. However, while it is easy to write down an action for the RW model, it is nontrivial for the AOM. In fact, the equation of motion (15) is third order in time derivatives, but the simplest form of the action principle only generates equations of motion of even derivative order. There is a trick to obtain differential equations of odd order that will show how the AOM includes an extra degree of freedom of non-associative character.

The action of the RW model (3) and of the DOM is (Calcagni et al., 2020)

$$S = \int_0^T dt \, e^{2\alpha\beta t}\left[\frac{\dot{v}^2}{2} - U(v)\right], \tag{16}$$

where $T$ is some final time and the potential $U(v)$ is given by (5) for the DOM and by (5) with $\mu = 0$ for the RW model. The action (16) only depends on one degree of freedom, the association strength $v(t)$. Varying $S$ with respect to $v(t)$ yields the equation of motion (7) or (9). However, it is not possible to find an action for the AOM only in terms of $v(t)$ and we must include a new degree of freedom, which is technically called a Lagrange multiplier and that we will dub it $y(t)$. Then, varying the action

$$S_{\text{AOM}} = \int_0^T dt \, y \left(\frac{d^2}{dt^2} + \mu^2\right)[\dot{v} + \alpha\beta(v - \lambda)] \tag{17}$$

with respect to $y$ immediately yields the equation of motion (15). The variation with respect to $v$ gives an independent equation for the extra degree of freedom,

$$\left(\frac{d^2}{dt^2} + \mu^2\right)(\dot{y} - \alpha\beta y) = 0 \,,$$

whose solution is

$$y(t) = e^{\alpha\beta\, t} + A \sin \mu t + B \cos \mu t \,. \tag{A4}$$

A key difference with respect to the profile (12) of the association strength is that $y(t)$ increases indefinitely in the future. One might interpret it as motivation, assuming no room for boredom effects in this model.

There are other ways to define an action for the AOM, but they entail more extra degrees of freedom.

**Appendix B – BIC and AIC tables**

Tables 3-6 show the BIC and AIC (presented in the format "BIC, AIC") of the best fits of the three models for all subjects of Harris et al. (2015) experiment. Calling the difference $\Delta_{12} = (\text{IC model 1}) - (\text{IC model 2})$ for the Bayesian or Akaike IC, one finds evidence in favor of model 2 if $\Delta_{12} > 0$. Following the classification of Jeffreys (1961) and Kass and Raftery (1995), this evidence is weak if $\Delta_{12} < 2$ (blue color in Tabs. 3-6), positive if $2 \leq \Delta_{12} < 6$ (green), strong if $6 \leq \Delta_{12} < 10$ (orange), and very strong if $\Delta_{12} \geq 10$ (red). In the table, all numbers are rounded to zero decimals, but to calculate $\Delta_{12}$ all digits were used. For each IC, "winners" with respect to the second best are in boldfaced color. The order of comparison is always RW – DOM ($\Delta_{RD}$) and AOM – RW or AOM – DOM ($\Delta_N$), depending on which between RW and the DOM is the best or second best in each IC (the comparison in a given IC is made with the second best model, unless this model is the third best in the other IC). When two models are favored in different ICs, the one with strongest evidence wins and the one winning on the other two is the winner (for instance, for CR10-8, the AOM is favored over RW but the DOM is favored over both). For subjects CR30-15 and PR10-10, where positive evidence is balanced between RW and the AOM, we declared RW the winner by Occam's razor. Trivial RW fits with vanishing $\alpha\beta$ are reported except when also the other fits fail, in which case all cells are left blank. All cases have a $p$ value $p \ll 0.001$ for nonzero $\mu$ except those indicated with an asterisk, which may be false positives since $p \approx 1$.

The results for the experiment of Calcagni et al. (2020) are given in Tabs. 7 and 8. Using the BIC instead of the AIC, Figs. 5 and 10 do not change appreciably. The overall effect is a global decrease of the evidence in favor of the DOM/AOM (all points are shifted downwards). Some of the points that were slightly above the zero line (weakly favored DOM/AOM with respect to RW) now lie below the line (weakly favored RW), but the great majority are still in the upper half and, most importantly, there is still the same strong evidence towards the DOM/AOM.

| Subject | RW | DOM | AOM | $\Delta_{RD}$ | $\Delta_N$ |
|---|---|---|---|---|---|
| **1** | 130, 120 | **55, 43** | 80, 60 | 74, 78 | 24, 18 |
| **2** | 177, 168 | 182, 170 | **171, 155** | −5, −2 | −6, −13 |
| **3** | **141**, 131 | 146, 133 | 147, **131** | −5, −2 | 6, −1 |
| **4** | 119, 109 | 124, 111 | **105, 86** | −5, −2 | −13, −23 |
| **5** | 99, 90 | **80, 64*** | 104, 84 | 19, 26 | −19, 20 |
| **6** | 32, 23 | **21, 9** | 26, 10 | 11, 14 | 5, 1 |
| **7** | −58, −68 | **−76, −88** | −54, −70 | 17, 21 | −17, −19 |
| **8** | **141**, 132 | 142, **130** | 145, **129** | −1, 2 | 3, 0 |
| **9** | **140**, 130 | 143, 130 | 145, **129** | −3, 0 | 6, −1 |
| **10** | **8**, −1 | 13, 1 | 11, **−5** | −5, −2 | 3, −4 |
| **11** | 161, 152 | 165, 153 | 161, **145** | 8, 15 | 8, 8 |
| **12** | **102**, 92 | 107, 94 | 107, **91** | −5, −2 | 5, −1 |
| **13** | 82, 72 | 87, 74 | **77, 61** | −5, −2 | −5, −11 |
| **14** | 50, 40 | **49, 36** | 59, 40 | 1, 4 | 10, 3 |
| **15** | 202, 193 | **193, 177** | 206, 190 | 9, 16 | 13, 13 |
| **16** | 144, 134 | 143, 130 | **143, 127** | 1, 5 | 0, −3 |

*Table 3: BIC and AIC for the RW model, DOM, and AOM best fits of subjects in Group CR10.*

| Subject | RW | DOM | AOM | $\Delta_{RD}$ | $\Delta_N$ |
|---|---|---|---|---|---|
| **1** | 149, 139 | 141, 128 | 138, 122 | 8, 11 | −2, −5 |
| **2** | 207, 197 | 211, 198 | 212, 196 | −4, −1 | 5, −2 |
| **3** | 23, 13 | 28, 15 | 30, 14 | −5, −2 | 7, 0 |
| **4** | 119, 109 | 117, 104 | | 2, 5 | |
| **5** | 115, 106 | 120, 108 | 26, 7 | 74, 80 | −16, −19 |
| **6** | | | | | |
| **7** | −101, −111 | −120, −133 | −111, −127 | 19, 23 | 9, 6 |
| **8** | 163, 154 | 168, 156 | 167, 151 | −5, −2 | 4, −2 |
| **9** | 57, 48 | 63, 50 | 53, 37 | −5, −2 | −5, −11 |
| **10** | −83, −92 | −78, −91 | −89, −105 | −4, −1 | −7, −13 |
| **11** | 103, 94 | 87, 74 | 92, 76 | 17, 20 | 6, 2 |
| **12** | 92, 82 | 71, 58 | 52, 33 | 21, 24 | −19, −26 |
| **13** | −36, −46 | −31, −44 | −31, −47 | −5, −2 | 5, −1 |
| **14** | −87, −96 | −94, −107 | −100, −119 | 8, 11 | −6, −12 |
| **15** | 96, 86 | 102, 86 | 99, 83 | −6, 0 | 3, −3 |
| **16** | 150, 141 | 156, 143 | 156, 140 | −5, −2 | 5, −1 |

*Table 4: BIC and AIC for the RW model, DOM, and AOM best fits of subjects in Group CR30.*

| Subject | RW | DOM | AOM | $\Delta_{RD}$ | $\Delta_N$ |
|---|---|---|---|---|---|
| **1** | 533, 520 | 450, 433 | 537, 515 | 83, 87 | 87, 82 |
| **2** | 458, 445 | 460, 443 | 468, 446 | −2, −2 | 10, 1 |
| **3** | 513, 500 | 519, 502 | 503, 482 | −6, −2 | −10, −19 |
| **4** | 514, 501 | 482, 465 | | 31, 36 | |
| **5** | 416, 403 | 184, 163 | 426, 404 | 231, 240 | 241, 241 |
| **6** | | | | | |
| **7** | 160, 147 | 166, 149 | 165, 144 | −6, −2 | 5, −3 |
| **8** | 483, 470 | 488, 471 | 468, 443* | −5, −1 | −14, −27 |
| **9** | 215, 203 | 222, 205 | 185, 159 | −6, −2 | −30, −43 |
| **10** | −75, −88 | −71, −88 | −71, −92 | −5, 0 | 4, −4 |
| **11** | | | | | |
| **12** | 391, 378 | 368, 351 | 394, 373 | 23, 28 | 26, 22 |
| **13** | 344, 331 | 350, 333 | 346, 325* | −6, −2 | 2, −7 |
| **14** | 25, 12 | 19, 2 | 26, 5 | 6, 10 | 7, 3 |
| **15** | 439, 426 | 445, 428 | 445, 423 | −6, −2 | 6, −3 |
| **16** | 581, 568 | 587, 570 | 588, 567 | −6, −2 | 7, −1 |

*Table 5: BIC and AIC for the RW model, DOM, and AOM best fits of subjects in Group PR10.*

| Subject | RW | DOM | AOM | $\Delta_{RD}$ | $\Delta_N$ |
|---|---|---|---|---|---|
| **1** | 301, 288 | 309, 287 | 219, 197 | −8, 1 | −82, −91 |
| **2** | 555, 542 | 533, 516 | | 22, 26 | |
| **3** | 334, 322 | 341, 324 | 344, 318 | −6, −2 | 9, −4 |
| **4** | 316, 303 | 310, 293 | | 6, 10 | |
| **5** | 129, 116 | 136, 118 | 136, 114 | −6, −2 | 6, −2 |
| **6** | −343, −356 | −350, −367 | | 7, 11 | |
| **7** | −199, −212 | −217, −234 | −204, −225 | 18, 22 | 13, 9 |
| **8** | 244, 231 | 230, 208 | 243, 221 | 14, 22 | 13, 13 |
| **9** | 109, 96 | 115, 98 | 106, 84 | −6, −2 | −3, −12 |
| **10** | −190, −202 | −184, −206 | −201, −222 | −5, 3 | −11, −20 |
| **11** | 211, 198 | 62, 40 | 219, 197 | 150, 158 | 157, 157 |
| **12** | 208, 196 | 202, 185 | 210, 189 | 7, 11 | 8, 4 |
| **13** | 28, 15 | 26, 9 | 29, 7 | 2, 6 | 3, −1 |
| **14** | −174, −187 | −201, −223 | −168, −190 | 27, 36 | 33, 33 |
| **15** | 311, 298 | 296, 275 | 305, 284 | 15, 24 | 9, 9 |
| **16** | 724, 711 | 724, 707 | 728, 706 | 0, 4 | 4, −1 |

*Table 6: BIC and AIC for the RW model, DOM, and AOM best fits of subjects in Group PR30.*

| Subject | RW | DOM | AOM | $\Delta_N$ |
|---|---|---|---|---|
| **1-1** | 171, 163 | 175, 165 | 176, 163 | 5, 0 |
| **1-2** | 43, 36 | 48, 38 | 50, 37 | 7, 2 |
| **1-3** | −32, −40 | −28, −38 | −26, −38 | 6, 1 |
| **1-4** | 68, 60 | 71, 61 | 73, 61 | 5, 0 |
| **1-5** | −62, −69 | −57, −67 | −59, −72 | 3, −2 |
| **1-6** | 13, 5 | 18, 5 | | |
| **1-7** | 116, 108 | 91, 78 | 102, 89 | 11, 11 |
| **1-8** | – | | | – |

*Table 7: BIC and AIC for the RW model, DOM, and AOM best fits of subjects in Group 1. Subject 8 was removed from the group due to poor health.*

| Subject | RW | DOM | AOM | $\Delta_N$ |
|---|---|---|---|---|
| **2-1** | 64, 56 | 65, 55 | 67, 54 | 3, −2 |
| **2-2** | 103, 95 | 107, 97 | 103, 90 | 1, −4 |
| **2-3** | −3, −10 | −4, −14 | −1, −14 | 2, 0 |
| **2-4** | −27, −35 | −33, −43 | −30, −42 | −2, −7 |
| **2-5** | 84, 76 | 85, 75 | 84, 71 | 0, −5 |
| **2-6** | 49, 41 | 52, 42 | 45, 32 | −4, −9 |
| **2-7** | −139, −147 | −135, −145 | −183, −196 | −44, −49 |
| **2-8** | −75, −82 | −76, −86 | −68, −80 | 8, 6 |

*Table 8: BIC and AIC for the RW model, DOM, and AOM best fits of subjects in Group 2.*

## References

Blanco, F., & Moris, J. (2018). Bayesian methods for addressing long-standing problems in associative learning: The case of PREE. *Q. J. Exper. Psychol. 71*, 1844.

Bush, R.R., & Mosteller, F. (1951a). A mathematical model for simple learning. *Psychol. Rev. 58*, 313; reprinted in Mosteller (2006), pp. 221–234.

Bush, R.R., & Mosteller, F. (1951b). A model for stimulus generalization and discrimination. Psychol. Rev. 58, 413; reprinted in Mosteller (2006), pp. 235–250.

Calcagni, G. (2018). The geometry of learning. *J. Math. Psychol. 84*, 74.

Calcagni, G., Caballero-Garrido, E. & Pellón, R, (2020). Behavior stability and individual differences in Pavlovian extended conditioning. *Frontiers Psychol. 11*, 612.

Estes,W.K. (1950) Toward a statistical theory of learning. *Psychol. Rev. 57*, 94.

Gallistel, C.R. (2012). On the evils of group averaging: Commentary on Nevin's "Resistance to extinction and behavioral momentum". *Behav. Proc. 90*, 98.

Gallistel, C.R., Fairhurst, S., & Balsam, P. (2004). The learning curve: implications of a quantitative analysis. *Proc. Nat. Acad. Sci. USA 101*, 13124.

Ghirlanda, S., & Enquist, M. (1998). Artificial neural networks as models of stimulus control. *Anim. Behav. 56*, 1383.

Ghirlanda, S., & Enquist, M. (2019). On the role of responses in Pavlovian acquisition. *J. Exp. Psychol.: Anim. Learn. Cogn. 45*, 59.

Ghirlanda, S., & Ibadullaiev, S. (2015). Solution of the comparator theory of associative learning. *Psychol. Rev. 122*, 242.

Glautier, S. (2013). Revisiting the learning curve (once again). *Front. Psychol. 4*, 982.

Goodfellow, I., Bengio, Y., & Courville, A. (2016). *Deep Learning*. Cambridge, MA: MIT Press, pp. 200–220.

Harris, J.A., Patterson, A.E., & Gharaei, S. (2015). Pavlovian conditioning and cumulative reinforcement rate. *J. Exp. Psychol. Anim. Learn. Cogn. 41*, 137.

Hayes, K.J. (1953). The backward curve: a method for the study of learning. *Psychol. Rev. 60*, 269.

Hull, C.L. (1943). *Principles of Behavior*. New York, NY: Apple-Century-Crofts.

Jaksic, H., Vause, T., Frijters, J.C., & Feldman, M. (2018). A comparison of a novel application of hierarchical linear modeling and nonparametric analysis for single-subject designs. *Behav. Anal. Res. Prac. 18*, 203.

Jeffreys, H. (1961). *Theory of Probability* (3rd ed.). Oxford University Press.

Kass, R.E., & Raftery, A.E. (1995). Bayes factors. *J. Amer. Stat. Assoc. 90*, 773.

Le Pelley, M.E. (2004). The role of associative history in models of associative learning: a selective review and a hybrid model. *Q. J. Exp. Psychol. 57*, 193.

Mackintosh, N.J. (1975). A theory of attention: Variations in the associability of stimuli with reinforcement. *Psychol. Rev. 82*, 276.

Mazur, J.E., & Hastie, R. (1978). Learning as accumulation: A reexamination of the learning curve. *Psychol. Bull. 85*, 1256.

McClelland, J.L., & Rumelhart, D.E. (Eds.) (1988). *Explorations in Parallel Distributed Processing: A Handbook of Models, Programs, and Exercises.* Cambridge, MA: MIT Press.

Merrill, M. (1931). The relationship of individual growth to average growth. *Hum. Biol. 3*, 37.

Miller, R.R., Barnet, R.C., & Grahame, N.J. (1995). Assessment of the Rescorla–Wagner model. *Psychol. Bull. 117*, 363.

Miller, R.R., Greco, C., & Vigorito, M. (1981). Classical conditioned tail flexion in rats: CR-contingent modification of US intensity as a test of the preparatory response hypothesis. *Anim. Learn. Behav. 9*, 80.

Mosteller, F. (2006). S.E. Fienberg & D.C. Hoaglin (Eds.), *Selected Papers of Frederick Mosteller*. New York, NY: Springer.

Pearce, J.M., & Hall, G. (1980). A model for Pavlovian learning: variations in the effectiveness of conditioned but not of unconditioned stimuli. *Psychol. Rev. 87*, 532.

Rescorla, R.A., & Wagner, A.R. (1972). A theory of Pavlovian conditioning: variations in the effectiveness of reinforcement and nonreinforcement. In A.H. Black & W.F. Prokasy (Eds.), *Classical Conditioning II* (pp. 64–99). New York, NY: Appleton-Century-Crofts.

Schultz, W., Dayan, P., & Montague, P.R. (1997). A neural substrate of prediction and reward. *Science, 275*, 1593.

Sidman, M. (1952). A note on functional relations obtained from group data. *Psychol. Bull. 49*, 263.

Smith, P.L., & Little, D.R. (2018). Small is beautiful: in defense of the small-N design. *Psychon. Bull. Rev. 25*, 2083.

Spence, K. (1956). *Behavior Theory and Conditioning*. New Haven, CT: Yale University Press.

Wagner, A.R. (1981). SOP: a model of automatic memory processing in animal behavior. In N.E. Spear & R.R. Miller (Eds.), *Information Processing in Animals: Memory Mechanisms* (pp. 5–47). Hillsdale, NJ: Erlbaum.

Wagner, A.R., & Rescorla, R.A. (1972). Inhibition in Pavlovian conditioning: applications of a theory. In M.S. Halliday & R.A. Boakes (Eds.), *Inhibition and Learning* (pp. 301–336). London, UK: Academic Press.

Wagner, A.R., & Vogel, E.H. (2009). Conditioning: theories. *Encyclopedia of Neuroscience 3*, 49.

Young, M.E. (2018). A place for statistics in behavior analysis. *Behav. Anal. Res. Prac. 18*, 193.

Zelikowsky, M., & Fanselow, M.S. (2010). Opioid regulation of Pavlovian overshadowing in fear conditioning. *Behav. Neurosci. 124*, 510.